\documentclass[useAMS,usenatbib,fleqn,usedcolumn]{mnras}

\usepackage[british]{babel}             
\usepackage{graphicx}                   

\usepackage{bm}
\usepackage{amsmath}
\usepackage{amssymb}

\newcommand{\unit}[1]{\ensuremath{\,\mathrm{#1}}}
\newcommand{\diff}{\ensuremath{\mathrm{d}}}
\newcommand{\deriv}[2]{\ensuremath{\frac{\diff #1}{\diff #2}}}
\newcommand{\pderiv}[2]{\ensuremath{\frac{\partial #1}{\partial #2}}}

\newcommand{\St}{\ensuremath{\mathrm{St}}}
\newcommand{\SigG}{\ensuremath{\Sigma_{\rm G}}}
\newcommand{\SigD}[1]{\ensuremath{\Sigma_{{\rm D} #1}}}

\newcommand{\reply}[1]{#1}

\title[Chemical enrichment]{Chemical enrichment of giant planets and discs due to pebble drift}
\author[Booth et al.]{Richard A. Booth\thanks{E-mail: rab200@ast.cam.ac.uk}$^1$, Cathie J. Clarke$^1$, Nikku Madhusudhan$^1$, and John D. Ilee$^1$ \\
$^{1}$Institute of Astronomy, University of Cambridge, Madingley Road, Cambridge, CB3 0HA, United Kingdom \\
}

\begin{document}

\date{Accepted 2017 May 4. Received 2017 May 4; in original form 2017 March 13}

\pagerange{\pageref{firstpage}--\pageref{lastpage}} \pubyear{2017}

\maketitle

\label{firstpage}

\begin{abstract}
Chemical compositions of giant planets provide a means to constrain how and where they form. Traditionally, super-stellar elemental abundances in giant planets were thought to be possible due to accretion of metal-rich solids. Such enrichments are accompanied by oxygen-rich compositions (i.e. C/O below the disc's value, assumed to be solar, $\mathrm{C/O} =0.54$). Without solid accretion the planets are expected to have sub-solar metallicity, but high C/O ratios. This arises because the solids are dominated by oxygen-rich species, e.g. H$_2$O and CO$_2$, which freeze out in the disk earlier than CO, leaving the gas metal poor but carbon-rich. Here we demonstrate that super-solar metallicities can be achieved by gas accretion alone when growth and radial drift of pebbles are considered in protoplanetary discs. Through this mechanism planets may simultaneously acquire super-solar metallicities and super-solar C/O ratios. This happens because the pebbles transport volatile species inward as they migrate through the disc, enriching the gas at snow lines where the volatiles sublimate. Furthermore, the planet's composition can be used to constrain where it formed. Since high C/H and C/O ratios cannot be created by accreting solids, it may be possible to distinguish between formation via pebble accretion and planetesimal accretion by the level of solid enrichment. Finally, we expect that Jupiter's C/O ratio should be near or above solar if its enhanced carbon abundance came through accreting metal rich gas. Thus Juno's measurement of Jupiter's C/O ratio should determine whether Jupiter accreted its metals from carbon rich gas or oxygen rich solids. 
\end{abstract}


\section{Introduction}
\label{Sec:Intro}

The chemical compositions of giant planets can provide insights into how they formed due to variations in the chemical composition of their host discs \citep{Oberg2011,Madhusudhan2014c,Cridland2016,Madhusudhan2016,Eistrup2016,Ali-Dib2017}. One key aspect is how and where they accreted their metals, because the abundance ratios such as C/N or C/O are correlated with their absolute abundances with respect to hydrogen in a way that depends on where the planet formed. In particular, because oxygen rich species such as H$_2$O and CO$_2$ freeze out at higher temperatures than CO, the dust grains in protoplanetary discs always have sub-solar C/O ratios, leaving the gas super-solar in C/O. Thus measurements of the C/O ratio provides a way of determining whether the planet accreted its metals predominantly by accreting gas, or solids in the form of dust grains or planetesimals. 

Since planet formation theories make differing predictions about when and how much solids are accreted, chemical constraints on the amount of solids accreted are able to place constraints on how the planets formed too. For example, giant planets that form through the gravitational instability are thought to form early \citep{Boss2000,Clarke2009,Rice2009,Boley2010,Forgan2011}. Since the initial accretion of gas and solids should be equally efficient for planets formed by the gravitational instability, they will acquire similar compositions to their host stars as their is little time for chemical processing in the disc. However, in core accretion giant planets must first grow 10 to 20$\unit{M_\oplus}$ rocky cores before accreting their gaseous envelopes. The classical model for core formation by runaway planetesimal collisions typically requires several Myr \citep{Pollack1996,Kobayashi2011}, which is long compared with the disc's lifetime \citep[a few Myr,][]{Haisch2001,Currie2009,Fedele2010,Meng2017}. Eventually the planets undergo run away accretion of gas and rapidly accrete their gaseous envelopes, but not until after significant chemical processing may have occurred in the disc \citep{Eistrup2016}.

Recently a variant of the standard core accretion model, known as pebble accretion, has been shown to overcome the growth time-scale problem, producing the cores of giant planets within $\sim 10^5\unit{yr}$ \citep{Ormel2010,Lambrechts2012}. In this model planetary embryos grow by accreting mm or cm-sized pebbles from a gas rich disc. Observations of protoplanetary discs show that there is a reservoir of such pebbles from which the embryos can accrete \citep{Wilner2005,Rodmann2006,Lommen2007,Perez2012,Tazzari2016}, making pebble accretion an attractive model for the formation of giant planets. 

One important difference between the standard planetesimal accretion scenario and pebble accretion is in the prospects for the late accretion of solids. Once the planetary cores become massive enough to perturb the gas surface density in the disc, the flux of pebbles from the outer disc is trapped in a pressure maximum outside the planet's orbit and pebble accretion stops \citep{Lambrechts2014a, Rosotti2016}. The planet then slowly accretes gas as the envelope contracts until eventually run-away gas accretion occurs. Without the presence of planetesimals, this results in a relatively clean envelope that has not been enriched by solids. Conversely, in the planetesimal accretion scenario the accretion of solids continues during the envelope contraction phase \citep{Pollack1996}, and may continue subsequently until the planet's feeding zone becomes depleted in planetesimals \citep[e.g][]{Alibert2005}, possibly leading to highly enriched atmospheres. We note that in practice there is likely to be some middle ground between these two extremes (e.g. the presence of some planetesimals even if the cores form by pebble accretion so it's not clear how much enhancement to expect). The enrichment of planetary atmospheres is further complicated by whether the planetesimals are disrupted in the envelope or reach the core \citep{Mordasini2015,Pinhas2016}, along with the possibility that the planet's core may be partially eroded and mixed throughout the envelope \citep{Wilson2012}.

With the robust detection of molecular species such as H$_2$O, CH$_4$, CO, and CO$_2$ becoming possible in the atmospheres of hot Jupiters, the idea of testing planet formation using planet composition is very appealing \citep{Lodders2002,Madhusudhan2012,Moses2013,Tsai2016}. In particular, the water abundance in the spectrum of a hot Jupiter provides a strong constraint on the C/O ratio: if the observed water abundance is far below that  expected for \reply{the stellar} C/O ratio, then one can infer the C/O ratio is close to unity \citep{Madhusudhan2012}, modulo any possible uncertainty due to obscuration of the water feature by clouds or hazes \citep[see][and references therein, for a recent discussion]{Madhusudhan2016}. Nevertheless, observations show that hot Jupiters may have C/O ratios that range from approximately \reply{stellar} \citep{Kreidberg2014,Line2014,Haynes2015} to ${\rm C/O} \gtrsim 1$ \citep{Madhusudhan2011,Stevenson2014}\footnote{\reply{Throughout this work we use the solar C/O ratio (${\approx 0.54}$) as our base value. As noted by \citet{Brewer2017}, when considering exoplanets the reader should bear in mind that the comparison must be made to the stellar abundances, which can be sub-solar.}}.

The solar system giant planets provide another opportunity to probe planet formation. Models can place constraints on their location during terrestrial planet formation, requiring them to be near their current locations during the process \citep[e.g.][]{Tsiganis2005,Raymond2006,Walsh2011,Agnor2012,Raymond2014}. This suggests that they either formed in situ, or migrated through the disc. Other `disc-free' modes of migration such as planet-planet scattering or the Kozai-Lidov mechanism \citep{Kozai1962,Lidov1962} can be ruled out, due to the giant planets' low eccentricities. Furthermore, the lack of a suitable companion and the long time required also disfavour the Kozai-Lidov mechanism. 

While we know that Jupiter is carbon rich, with a C/H ratio approximately 3 to 5 times solar, and is also enriched in noble gases, along with nitrogen and sulphur \citep{Owen1999,Atreya2005}, currently the oxygen abundance is uncertain. Saturn's abundances are in general more uncertain, but the C/H ratio is well determined to be about 8 times solar and the N/H ratio is around 3 time solar \citep{Atreya2016}. Although the oxygen abundance remains uncertain for both Jupiter and Saturn, the Galileo probe's measurement of water in a hot spot in Jupiter's atmosphere provides a possible constraint \citep{Owen1999,Atreya2005,Atreya2016}. Galileo found an O/H ratio of about one half solar, giving a ${\rm C/O \approx 10}$, but it is not clear that the measurement is representative of Jupiter's bulk atmosphere. Using thermochemical models, \citep{Visscher2010} were able to place a lower limit on the C/O ratio, suggesting that C/O ratios that are less than about half solar ($\mathrm{C/O} \lesssim 0.25$) are incompatible with the levels of CO detected in Jupiter's atmosphere. However, the Juno satellite should soon provide us with a reliable measurement of water in Jupiter \citep{Matousek2007}, which can be used to determine the global O/H ratio \citep{Helled2014}. The imminent availability of Juno's measurement of Jupiter's water abundance provides us with an excellent opportunity to constrain the formation of giant planets.

A super-solar C/O ratio along with a high C/H for Jupiter is in tension with current models of planet formation, which generally predict that planets with super-solar C/H ratios should normally have sub-solar C/O ratios \citep{Oberg2011, Madhusudhan2014c,Cridland2016,Madhusudhan2016,Eistrup2016,Ali-Dib2017}. The sub-solar C/O ratios arise because the additional carbon is brought into the planet's envelope by accreting \emph{solids} (either in the form of planetesimals, or by partially eroding the core) because the condensation of volatile species results in a depleted gas phase carbon abundance. Since H$_2$O and CO$_2$ condense at higher temperatures than CO, the solids are always oxygen rich unless a high fraction ($\sim 50$ per cent) of the carbon is locked up into refractory organic molecules. Similarly, the gas ends up with a sub-solar C/H ratio but super-solar C/O ratio. Combining varying amount of gas and solids results in planetary compositions that have either a C/H ratio above solar but a C/O ratio below solar, or a C/O ratio above solar with a C/H ratio below solar. Thus the compositions of giant planets are restricted to two octants of the C/H vs. O/H ratio plane \citep[see e.g.][]{Madhusudhan2014c,Madhusudhan2016}.

The above correlations between predicted elemental ratios arise naturally from partitioning solar composition gas into solids and ices according to condensation temperature of different molecular species. If the C/O ratio in Jupiter turns out to be high then this can be achieved in one of two ways: 1) change the C/O ratio of the solids accreted, for example by accreting from an as of yet unknown family of organic rich planetesimals \citep{Lodders2004} or 2) change the abundance of the gas to make it more carbon rich, such as through photoevaporation \citep{Guillot2006,Monga2015}.

In this paper we explore idea 2) and argue that high gas phase C/H and C/O ratios might arise naturally as a consequence of disc evolution. The key point is that the large reservoir of mm or cm-sized grains are expected to migrate radially. As the grains migrate they will carry any volatiles frozen onto their surface to snow lines, where the volatiles can desorb enhancing the local chemical abundance. Subsequent advection with the gas and diffusion may then spread these molecules inwards throughout the disc. This idea has been explored in relation to the water snow line or generic volatiles \citep{Stepinski1997,Hueso2003,Ciesla2006,Garaud2007}; however, increasing the abundance of water inside the water snow line is clearly of little help in enhancing the C/O ratio. However, the enhancement of CO can increase the CO ratio from its solar value of $\mathrm{C/O} \approx0.54$ \citep{Asplund2009} to nearly twice the solar value, along with the C/H ratio, as suggested by \citet{Oberg2016}.

The accretion of these metal rich gases can then directly produce metal rich planetary envelopes, without the need for subsequent accretion of solids or erosion of the core. Although this way of enhancing the envelope's abundances is independent of the way in which the planet's core is formed, we note that it fits naturally into the pebble accretion model because the pebbles responsible for the rapid accretion are also those that dominate the flux of volatiles. For this reason we have focussed on the pebble accretion model of planet formation, but giant planets with cores that formed by the accretion of planetesimals may also accrete metal rich gas this way. Indeed, the differences between the chemical abundances produced in our models and those discussed above are more due to the disc evolution model than the planet formation model \emph{per se}.

In \autoref{Sec:Model} we describe the global models that we use for the evolution of the gas and grain populations in the disc, for the formation of planets and for disc chemistry. In \autoref{Sec:Results:DiscEvo} we describe the resulting evolution of the disc, focusing on the evolution of elemental abundances at different locations in the disc. In \autoref{Sec:Results:PlanetForm} the results of the planet formation model are presented, along with the chemical composition of their atmospheres in \autoref{Sec:Results:PlanetChem}. In \autoref{Sec:Results:Jupiter} we discuss our models in the context of the formation of Jupiter and Saturn. In \autoref{Sec:Discuss} we discuss our results in terms of observations of protplanetary discs and planetary atmospheres. In \autoref{Sec:Conclude} we present our conclusions.

\section{Model}
\label{Sec:Model}

We base our model for dust evolution on the two-population model of \citet{Birnstiel2012}, which is ideal for our purposes because it has been calibrated to match the grain size and mass flux in detailed simulations of growth and radial drift. The evolution of the gas is treated using a simple viscous evolution model. The dust is treated by considering two populations of grains: small grains which remain tightly coupled to the gas and large grains which may grow and drift. We have extended the model to include the effects of feedback from the dust on the gas, which can become important in regions of high dust density. Additionally we follow evolution of the dominant carbon and oxygen bearing species, which may be present as varying fractions of vapour and ice. Our treatment for adsorption and desorption of molecular species follows insights from grain growth and chemical models, and agrees well with a detailed study on the delivery of CO to the CO snow line by \citet{Stammler2017}.

\subsection{Disc evolution}
\label{Sec:DiscEvo}

We model the disc evolution, following the viscous evolution of the gas, growth and radial drift of solids, and a simple model for the adsorption and desorption of the dominant carbon and oxygen bearing species. The coupled evolution of the dust and gas is solved in a `one-fluid' manner, following \citet{Laibe2014}. This approach involves solving for the evolution of the total surface density, 
\begin{equation}
 \Sigma = \SigG + \sum_i \SigD{i},
\end{equation}
along with the dust fractions $\epsilon_i = \SigD{i} / \Sigma$. Here $\SigG$ and $\SigD{i}$ are the surface density of the gas and the $i$th dust species. From conservation of mass and momentum, the usual viscous evolution equation \citep{Lynden-Bell1974} can be extended to the multi-fluid case,
\begin{equation}
 \pderiv{\Sigma}{t} = \frac{1}{R} \pderiv{}{R}\left[ 3 R^{1/2} \pderiv{}{R}\left(\nu \SigG R^{1/2}\right)\right] \label{Eqn:ViscEvo}
\end{equation}
where $\nu$ is the kinematic viscosity and we have made the usual approximation that the velocity profile does not significantly differ from Keplerian rotation. Additionally we assume that viscosity only drives angular momentum transport in the gas. We use an $\alpha$ model for the viscosity $\nu = \alpha c_s H$, where $c_s$ is the sound speed and $H = c_s / \Omega$ is the disc scale-height. The sound speed is calculated from the temperature assuming a mean-molecular weight, $\mu = 2.4$. To compute the temperature, we follow \citet{Birnstiel2010} and use the model of \citet{Hueso2005}, which includes viscous heating, irradiation from a central star and external radiation. For the opacity we use the updated \citet{Bell1994} type tabulated opacity provided by \citet{Zhu2012}. For the stellar properties we use ${M=1\,M_\odot}$, ${R=2.5\,R_\odot}$ and ${T_{\rm eff} = 4200\unit{K}}$, while we assume that external irradiation limits the temperature to be at least $10\unit{K}$.

By noting that the viscous radial velocity, $v_R$, is given by
\begin{equation}
\Sigma v_R = 3 R^{-1/2} \pderiv{}{R}\left(\nu \SigG R^{1/2}\right),
\end{equation}
a trace quantity with concentration, $\phi$, can be followed by advecting the mass of the tracer $\Sigma \phi$ at velocity $v_R$, 
\begin{equation}
\pderiv{(\Sigma \phi)}{t} = \frac{1}{R}\pderiv{}{R}\left[R  v_R (\Sigma \phi) \right]. \label{Eqn:ViscTracer}
\end{equation}
Numerically, we solve equations \ref{Eqn:ViscEvo} \& \ref{Eqn:ViscTracer} using a first order explicit finite-volume update following \citet{Bath1981}.

We follow \citet{Laibe2014} for the update of the dust fraction, who give an expression the evolution of $\epsilon_i$ in a Lagrangian frame, which we write as
\begin{equation}
 \deriv{\epsilon_i}{t} = \frac{1}{\Sigma R} \pderiv{}{R} 
 \left[ 
 R \Sigma \epsilon_i (\Delta v_i - \epsilon \Delta v) 
\right ]. \label{Eqn:DustUpdate}
\end{equation}
Here $\Delta v_i = v_i - v_g$, the difference between the dust and gas radial velocity and $\epsilon \Delta v = \sum_i \epsilon_i \Delta v_i$. To compute the relative velocity we follow \citet{Tanaka2005}, who provide an expression for the dust-gas relative velocity of a distribution of grain sizes when feedback is included. An equivalent form is given by \citet{Bai2010}. The radial component of the relative velocity is given by
\begin{equation}
\Delta v_i = \frac{2 v_\phi \St_i - v_r \St_i^2}{1 + \St_i^2},
\label{Eqn:DeltaV}
\end{equation}
where $v_r$ and $v_\phi$ are the radial and azimuthal velocity of the gas,
\begin{align}
v_r    =&  -        \frac{1}{\rho_{\rm G}\Omega}\deriv{P}{R} \frac{ \lambda_1 }{(1 + \lambda_0)^2 + \lambda_1^2}, \\
v_\phi =& \frac{1}{2}\frac{1}{\rho_{\rm G}\Omega}\deriv{P}{R} \frac{1+\lambda_0}{(1 + \lambda_0)^2 + \lambda_1^2}.
\end{align}
Here, the mid-plane values of the pressure, $P$, and gas density, $\rho_{\rm G}$ are used and
\begin{equation}
\lambda_k = \sum_i \left(\frac{\epsilon_i}{1 - \epsilon}\right)\left(\frac{\St_i^k}{1 + \St_i^2}\right).
\end{equation}
The Stokes number, $\St_i = \frac{\upi}{2} \rho_s s_i /\SigG$, i.e. assuming Epstein drag for spherical grains in the disc mid-plane, where $\rho_s = 1\unit{g\,cm}^{-3}$ is the internal density of the grain and $s_i$ is its radius. In the complete update of $\epsilon_i$ we first advect $\epsilon_i$ as a tracer in the viscous evolution step. Then we compute $\Delta v_i$ using centred finite differences and use an explicit upwind finite volume update for \autoref{Eqn:DustUpdate}. Trace species that move with the dust, such as ices, are advected analogously to those moving with the gas. 

Additionally, we include the diffusion of dust and trace species given by:
\begin{equation}
 {\pderiv{\phi}{t}} = \frac{1}{R\Sigma} \pderiv{}{R}\left[R D_\phi \Sigma \pderiv{\phi}{R}\right],
\label{Eqn:Diffuse}
\end{equation}
where $D_\phi = \nu / Sc_\phi$, where $Sc_\phi$ is the Schmidt number of the tracer. For gas tracers we take $Sc_\phi = 1$, while for dust we use $\phi_i = \epsilon_i$ and
\begin{equation}
Sc_i = \frac{(1 + \St_i^2)^2}{1 + 4 \St_i^2}
\end{equation}
\citep{Youdin2007}. This diffusive flux is included as an extra term in \autoref{Eqn:DustUpdate}.

\subsection{Grain growth}
\label{Sec:GrainGrowth}

For grain growth, we follow the prescription given in Appendix B of \citet{Birnstiel2012}. In this model, the large grains grow on a time-scale, $t_{\rm grow} = (\epsilon \Omega)^{-1}$, until their growth is prevented by radial drift or fragmentation. In the outer parts of the disc grain growth is limited as the radial drift time-scale becomes shorter than the growth time-scale and thus the grains are removed before they can grow. This occurs at a Stokes number,
\begin{equation}
\St_\mathrm{drift} = 0.55 \epsilon \left( \frac{R}{H} \right)^2 \left|\deriv{\ln P}{\ln R}\right|^{-1}. \label{Eqn:DriftLimit}
\end{equation}
At smaller radii growth is eventually limited by fragmentation driven by the turbulent velocities of dust grains, limiting their size to
\begin{equation}
\St_\mathrm{frag} = 0.37 \frac{1}{3\alpha}\frac{v_f^2}{c_s^2}, \label{Eqn:FragLimit}
\end{equation}
where $v_f$ is the threshold velocity for fragmentation.

The grains' initial size is taken to be $0.1\unit{\umu m}$. The mass fraction of grains in the large and small populations depends on whether their size is limited by fragmentation or radial drift: if fragmentation limits their growth then 75 per cent of the mass is in large grains, while the fraction of large grains is 97 per cent if radial drift dominates. 

Fragmentation occurs when grains collide at velocity above their fragmentation threshold, $v_f$. For icy grains, we take $v_f = 10\unit{m\,s}^{-1}$ \citep{Gundlach2015}, while for `dry' grains we use the fragmentation threshold for silicates, $v_f = 1\unit{m\,s}^{-1}$ \citep{Blum2008}. To determine whether grains are icy or ice-free, we compare the total mass in all of the ices on the grains to the mass of the grains themselves. If this is above 10 per cent we use the icy fragmentation threshold. For ice concentrations below 10 per cent we interpolate between the upper and lower thresholds. This choice ensures that the transition between icy and ice-free grains follows the water snow line as the disc evolves.

\subsection{Chemistry}
\label{Sec:Chemistry}
 \begin{table*}
  \centering
 \begin{tabular}{cccc}
  \hline \hline
  Species (X) & T$_{\rm bind}$ [K]$^a$ & Case 1$^b$: X/H & Case 2$^c$: X/H \\ \hline
  CO     &  850 & 0.45 ($1 + f_{{\rm CO}_2}$) $\times$ C/H   & 0.65 $\times$ C/H \\
  CH$_4$ & 1300 & 0.45 ($1 - f_{{\rm CO}_2}$) $\times$ C/H   & 0 \\
  CO$_2$ & 2575 & 0.1  $\times$ C/H                                    & 0.15 $\times$ C/H \\
  H$_2$O & 5700 & O/H - (3 $\times$ Si/H + CO/H + 2 $\times$ CO$_2$/H) & O/H - (3 $\times$ Si/H + CO/H + 2 $\times$ CO$_2$/H) \\
  Carbon grains & -- & 0 & 0.2 $\times$ C/H \\
  Silicates     & -- & Si/H & Si/H \\
  \hline
  \end{tabular}
  \caption{Binding energies and volume mixing ratios of chemical species in the model disk \citep[adapted from][]{Madhusudhan2014c,Madhusudhan2016}.
  }
  \label{Tab:Chem}
  {$^a$ The binding energies are adopted from \citet{Collings2004,Garrod2006} and \citet{Fayolle2016}, and produce ice lines at temperatures of $20\unit{K}$, $71\unit{K}$ \& $180\unit{K}$ for CO, CO$_2$ and H$_2$O respectively. We use the binding energy of pure CO, rather than the higher binding energy of a mixture of CO and water ice. This  produces a freeze out temperature closer to the $20\unit{K}$ used previously. \\
  $^{b,c}$ Volume mixing ratios (i.e. by number) adopted for the species as a function of disk elemental abundances under two different prescriptions for condensate chemistry \citep[see e.g.][]{Madhusudhan2014c}. Solar values are assumed for the elemental abundances C/H, O/H, and Si/H \citep{Asplund2009}. The Case 2 chemistry is adopted from \citet{Oberg2011} and contains carbon grains leading to more solid carbon. The fraction of CO$_2$ molecules in ices is denoted $f_{{\rm CO}_2}$.}
 \end{table*}

We use the simple chemical models of \citet{Madhusudhan2014c}, which include only the dominant carbon and oxygen bearing volatile species, H$_2$O, CO, CO$_2$, CH$_4$ \citep{Oberg2011, Madhusudhan2014c}. In \citet{Madhusudhan2014c} and \citet{Madhusudhan2016} two different models for the abundance fractions of these species were used, Cases 1 \& 2, shown in \autoref{Tab:Chem}. These abundance fractions are motivated by either theoretical chemical calculations (Case 1, \citealt{Woitke2009}) or observation of ice grains in protplanetary discs (Case 2, \citealt{Draine2003,Pontoppidan2006,Oberg2011}). 

In previous studies \citep{Madhusudhan2014c,Madhusudhan2016}, the molecular species were assumed to be entirely in either the gas phase or in ices on the surface of grains, depending on whether the local temperature exceeded some typical condensation temperature. However, such a sharp transition in the chemical abundance results in numerical difficulties because the diffusion of chemical species depends on their concentration gradient (\autoref{Eqn:Diffuse}), which would not be well defined. Instead, we solve for the equilibrium ice abundance by balancing the thermal adsorption and desorption rates in the mid plane. 

Following \citet{Hollenbach2009} and \citet{Visser2009} the thermal desorption rate \emph{per grain} is given by $4 \upi s_0^2 N_{s,i}R_{d,i} f_{s,i}$, where $N_{s,i}=1.5\times 10^{15}\unit{cm}^{-2}$ is the number of binding sites per unit area on the grain, where we have assumed spherical grains and $s_0$ is their mean size by area. $f_{s,i}$ is the fraction of binding sites covered by the molecule. The desorption rate per molecule is 
\begin{equation}
R_{d,i} = 1.6 \times 10^{11} \sqrt{\frac{T_{{\rm bind},i}}{\mu_i}} \exp \left( -T_{{\rm bind},i} / T \right),
\end{equation}
where $T_{{\rm bind},i}$ and $\mu_i$ are the binding energy and mean-molecular weight of the molecule. The binding energies used are given in \autoref{Tab:Chem}. For thermal adsorption we assume that all molecule-grain collisions result in sticking. Thus the adsorption rate per grain is given by $n_i \upi s_0^2 v_i $, where $n_i$ is the gas-phase number density of the molecule and $v_i$ is its thermal velocity. For the mean grain area we assume $0.1\,\umu m$ spherical grains. We note that although the dust mass is dominated by large grains, the surface area of the grains is dominated by small grains for both protoplanetary disc dust populations \citep{Birnstiel2011} and the interstellar medium distribution of grain sizes \citep{Mathis1977}\footnote{Although it may appear surprising, using simulations including the full grain size distribution \citet{Stammler2017} showed that the adsorption onto small grains is indeed the dominant contribution to the adsorption rate of molecules but that the large grains contain the most of the mass in ices and dominate the radial flux of ices. Our simplified model captures these behaviours.}. 

Grain creation or destruction processes have been neglected in the models, i.e. although the grain size may change, the amount of silicate or carbon in grains only changes due to advection and diffusion. The ice mass has been included in $\epsilon_i$ when computing the growth and dynamics. We have also compared our equilibrium adsorption-desorption model to models in which we treat these adsorption and desorption processes in a time-dependent manner and find no significant differences, because typically the adsorption time-scale is shorter than the radial drift time-scale, except in the extreme outer parts of the disc.

In all models we use one of the abundance distributions given in \autoref{Tab:Chem} for the initial conditions. However, as the disc evolves the abundances of the molecular species will change due to radial drift. These changes will drive the chemical abundances out of equilibrium and thus we might expect that subsequently chemical reactions would further modify the molecular abundances. We take a simplistic approach and try to capture the important behaviour instead of attempting to model all of the required chemical processes in detail, which would obscure our main focus of investigating the redistribution of molecular species. One way of achieving this is to view the abundances in \autoref{Tab:Chem} as models of the equilibrium chemical abundances, which chemical processes act to restore. We note that the equilibrium molecular abundances are likely to be dependent on the location in the disc, but for simplicity we neglect any such differences beyond those already included in \autoref{Tab:Chem}. 

We consider models that should bracket the possible behaviours by considering two scenarios: one in which reactions are assumed to be negligibly slow and are neglected and one in which the `equilibrium' abundances of \autoref{Tab:Chem} are immediately restored. In the first model the type of molecule in which each atom is contained is always remains the same, though it may adsorb, desorb, advect and diffuse through the disc, thus changing the local abundances. In the second model the molecules' abundances will additionally change through the conversion of one species into another close to the snow lines. We update the abundances of both the gas and ice phase molecular species, but assume that the amount of material locked up in carbon grains or silicates does not change. We note that due to strong enhancements of CO  near its snow line the conversion of CO to CO$_2$ may require more oxygen than is available. If this is the case then we create as much CO$_2$ as possible based on the available oxygen, leave the reaming excess carbon as CO and set the H$_2$O abundance to zero. 

For the Case 1 chemistry we have only considered the equilibrium abundance model since the CO and CH$_4$ abundance changes at the CO$_2$ snow line. Hence it is not clear that a slow reaction model makes sense for this chemical abundance pattern. Furthermore, since the fraction of CO$_2$ in ices now varies smoothly across the snow line we replace the sharp transition in the CO and CH$_4$ abundances with one smoothed by interpolating between the two regimes based upon the fraction of CO$_2$ in the gas phase. 

In reality, the effect of chemical reactions will lie between these two extremes of our models. \citet{Eistrup2016} show that the time-scale for the formation and destruction of the molecular species we consider typically occurs on time-scales of around a few $10^5\unit{yr}$. This time-scale falls between the radial drift time-scale (a few $10^4\unit{yr}$) and the time-scale for viscous evolution and gas diffusion ($\sim \mathrm{Myr}$). We can have some confidence that our models will capture the qualitative behaviour because the detailed chemical models of \citet{Eistrup2016} produce similar abundance patterns to our Case 1 chemical model when starting from the molecular abundances inherited from the parent cloud.

\begin{figure*}
 \includegraphics[width=\textwidth]{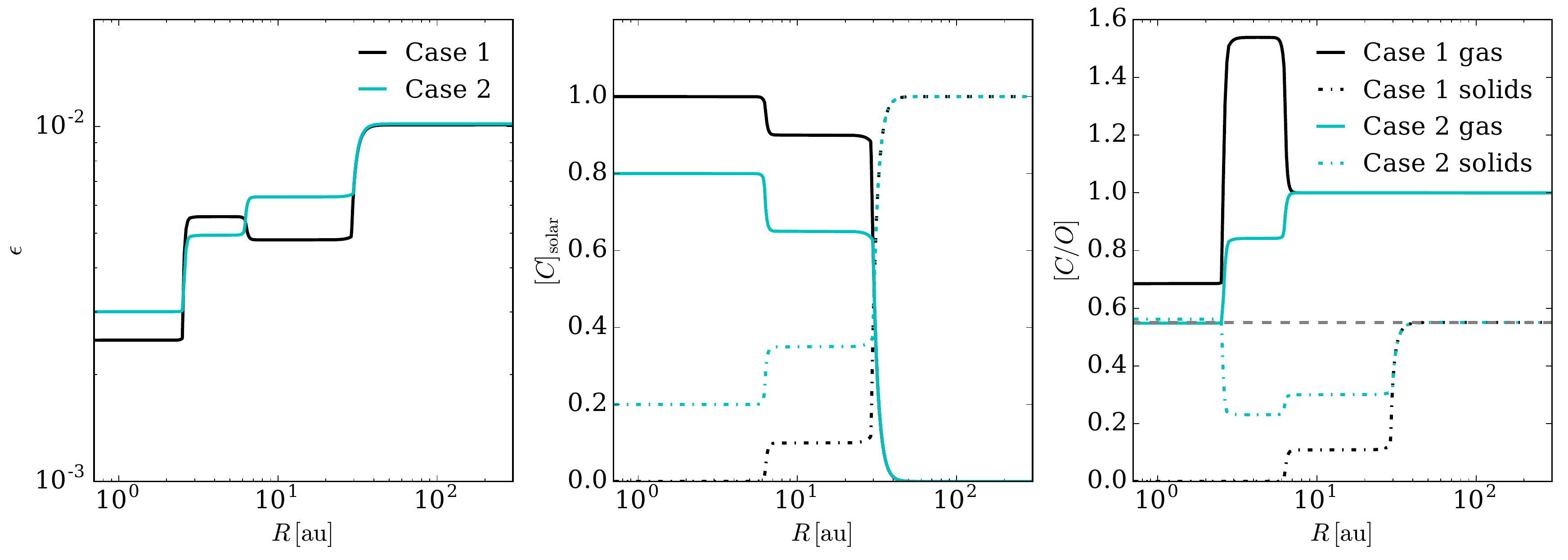}
\caption{Initial dust to gas ratio (left), carbon abundance (centre) and C/O ratio (right) for the two chemical models. The grey dashed line in the right hand panel shows the solar C/O ratio. The C/O ratio in the gas phase is always super-solar, but the gas phase carbon abundance is sub-solar. Note the carbon abundance in the middle panel is normalized by the total dust + gas mass for both the gaseous and solid carbon.}
\label{Fig:ICs}
\end{figure*}

\subsection{Planet formation}
\label{Sec:PlanetFormation}

We focus on giant planet formation by pebble accretion. For this we use the model based on  \citet{Bitsch2015}, which we only describe briefly. The only significant difference between our planetary growth model and the \citet{Bitsch2015} model is the way that we determine the surface density of and size of the pebbles used in the pebble accretion model. We use directly the values obtained from our disc evolution models, while \citet{Bitsch2015} use the prescriptions of \citet{Lambrechts2014b}. However, since the \citet{Lambrechts2014b} model is also based upon insights from grain growth models, such as the \citet{Birnstiel2012} model that we use, we do not expect the differences to be too great. 

Pebble accretion stops once the planet is massive enough to open a gap in the pebbles, which occurs when 
\begin{equation}
M_p \approx 20 \left(\frac{H/R}{0.05}\right)^3 M_\oplus
\end{equation}
\citep{Lambrechts2014a, Rosotti2016}. Initially, the gas accretion rate is slow and is controlled by envelope contraction. During this phase we follow \citet{Piso2014} and the planet's envelope grows according to
\begin{align}
\dot{M}_{\rm env}& = 1.75 \times 10^{-3} f^{-2}
	\left(\frac{\kappa_{\rm env}}{1\unit{cm}^{2}\unit{g}^{-1}}\right)^{-1}
	\left(\frac{\rho_{\rm c}}    {5.5\unit{g}\unit{cm}^{-3}}  \right)^{-1/6} \nonumber \\ 
        &\times\left(\frac{M_{\rm c}}       {M_\oplus}                   \right)^{11/3}
	\left(\frac{M_{\rm env}}     {0.1M_\oplus}                \right)^{-1}
	\left(\frac{T}               {81\unit{K}}                 \right)^{-1/2}
	\,M_\oplus\unit{Myr}^{-1}, \label{Eqn:Piso14}
\end{align}
where $M_{\rm c}$ is the core mass, $\kappa_{\rm env} = 0.05\unit{cm}^2\unit{g}^{-1}$ is the opacity of the envelope, $\rho_{\rm c} = {5.5\unit{g}\unit{cm}^{-3}}$ is the core's density and the factor $f=0.2$ \citep[see][for a discussion of these parameters]{Bitsch2015}. The planet's accretion rate during the envelope contraction only depends on the disc's properties through its local temperature, $T$. Once the envelope becomes as massive as the core, run away gas accretion occurs. We use the accretion rate from \citet{Machida2010}, with a maximum gas accretion rate limited to 80 per cent of the disc accretion rate. 

We have not included any late-phase pebble or planetesimal accretion, which would further enrich the envelope \citep{Mordasini2015,Pinhas2016}. We follow the chemical composition of the planets by tracking separately the mass accreted in solids and gas, along with the composition of those components. In this way we build up the total composition of the core and envelope of the planet. Subsequent erosion of the core may mix some of the solids into the envelope.

Planetary migration is also included following \citet{Bitsch2015}. For Type I migration we use the torque formulae of \citet*{Paardekooper2011}, while for Type II migration use the results of \citet{Baruteau2014}. We interpolate between these limits using the gap opening criterion of \citet{Crida2007}. Once the planets migrate close to the inner edge of the simulation we stop tracking their growth and migration.

We note the feedback of the planets on the disc has not been included: we have neglected the changes in disc mass due to planetary accretion or the surface density due to the torque from the planet.

\begin{figure*}
 \includegraphics[width=\textwidth]{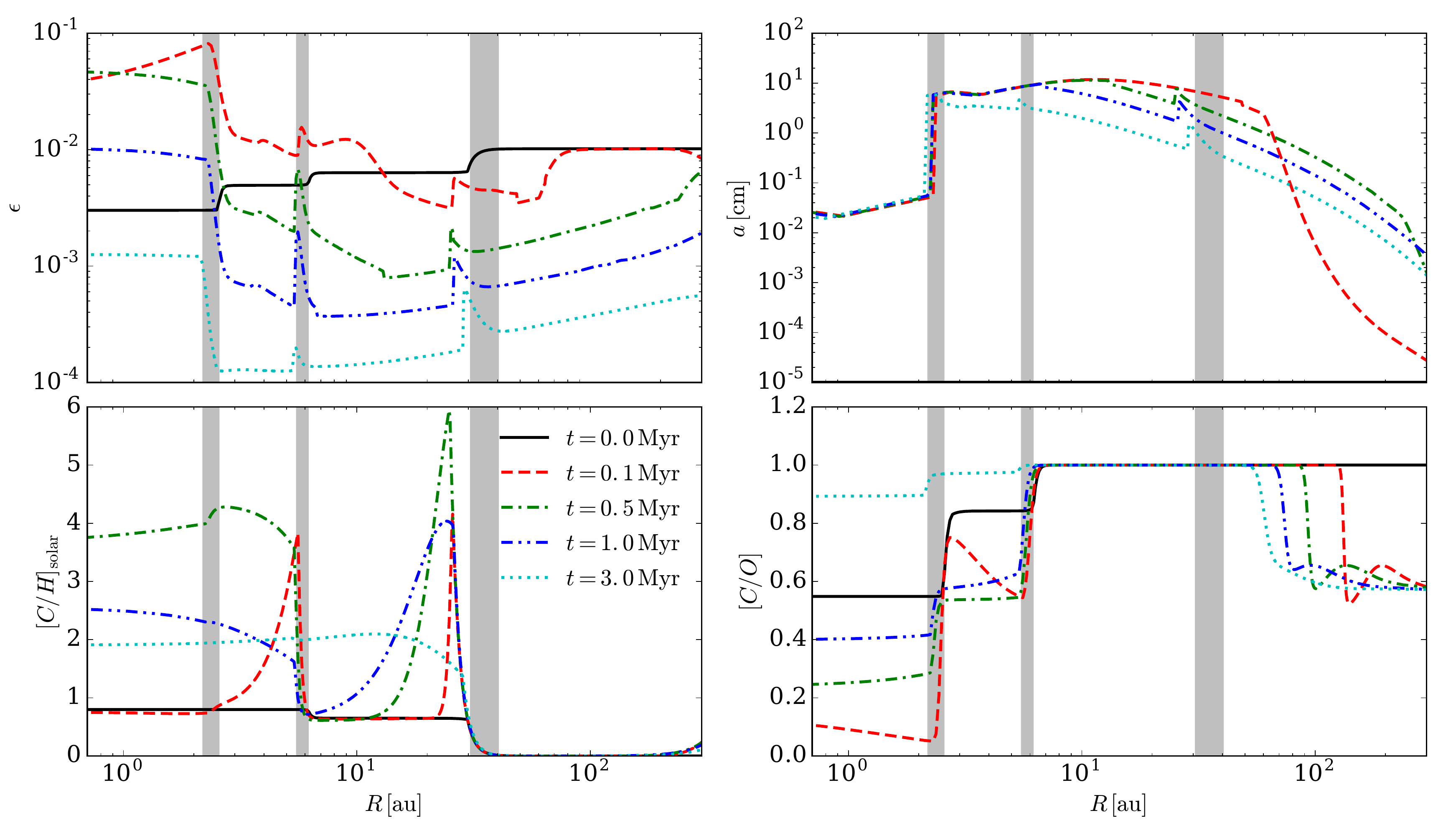}
\caption{Evolution of the dust fraction (top left), grain size (top right), gas phase carbon abundance relative to the solar abundance (bottom left) and gas phase C/O ratio (bottom right) for the Case 2 chemistry model. The jump in grain size at the water snow line (located at a few au) is due to the lower threshold velocity for fragmentation for non-icy grains. The grey bars show the regions where the H$_2$O, CO$_2$ and CO snow lines are located (from left to right, respectively).}
\label{Fig:ObergDiscTD}
\end{figure*}

\subsection{Initial conditions}
\label{Sec:ICs}

We set up the disc with an the initial surface density given by
\begin{equation}
\Sigma(R) = \frac{\dot{M}}{3\upi \nu} \exp \left( - R / R_d \right),
\end{equation}
and we typically use $\dot{M} = 10^{-8}\,M_\odot\unit{yr}^{-1}$,  $R_d = 100$ and $\alpha = 10^{-3}$ unless otherwise stated. Since $\nu$ depends on $\Sigma$ because the temperature depends on the optical depth, we iterate the initial surface density until $\nu$ and $\Sigma$ converge. We assume that the atomic abundances are initially solar \citep{Asplund2009} and the molecular abundances are given by either the Case 1 or Case 2 values. The dust mass is set to the sum of the carbon grains, silicate grains and any molecules that have condensed into ices according to \autoref{Tab:Chem}. In \autoref{Fig:ICs} we show the initial dust fraction and abundances for the two models. This figure shows clearly why models of planet formation without disc evolution predict that Jupiter must have a C/O ratio that is at most solar: Jupiter's high C/H ratio means that these model planets must have accreted their additional carbon in the form of solids, which are typically oxygen rich.

\section{Chemical evolution of the disc}
\label{Sec:Results:DiscEvo}
\subsection{No or negligibly slow reactions}

We first consider the evolution of the disc and chemistry for the model in which the chemical reactions are assumed to be slow, using the Case 2 abundances. Initially, the total gas+ice composition of each molecule is the same everywhere. \autoref{Fig:ObergDiscTD} shows the evolution of the fiducial disc model. The growth and radial drift of the grains causes the outer disc to become depleted of dust after about a Myr. Beyond $10\unit{au}$ the size of the grains is limited by radial drift, thus the maximum size of the dust grains also decreases due to the slower growth rate at lower dust-to-gas ratios. In the inner regions fragmentation limits grain growth and the maximum grain size does not evolve. The transition from icy to bare grains at the water snow line reduces the fragmentation threshold, resulting in smaller grains, lower radial drift speeds and higher dust fractions.

\begin{figure*}
\begin{tabular}{cc}
 \includegraphics[width=\columnwidth]{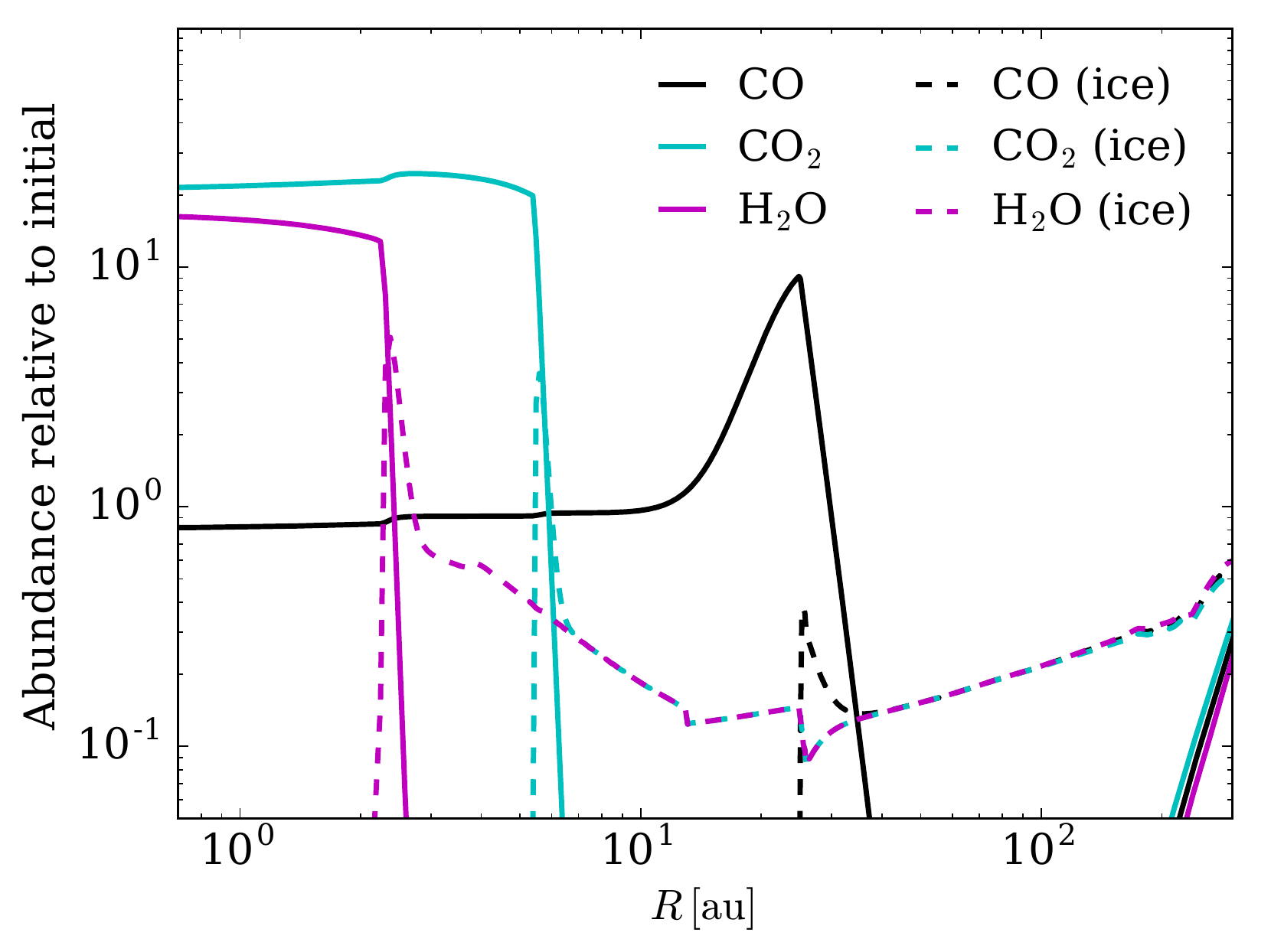} &
 \includegraphics[width=\columnwidth]{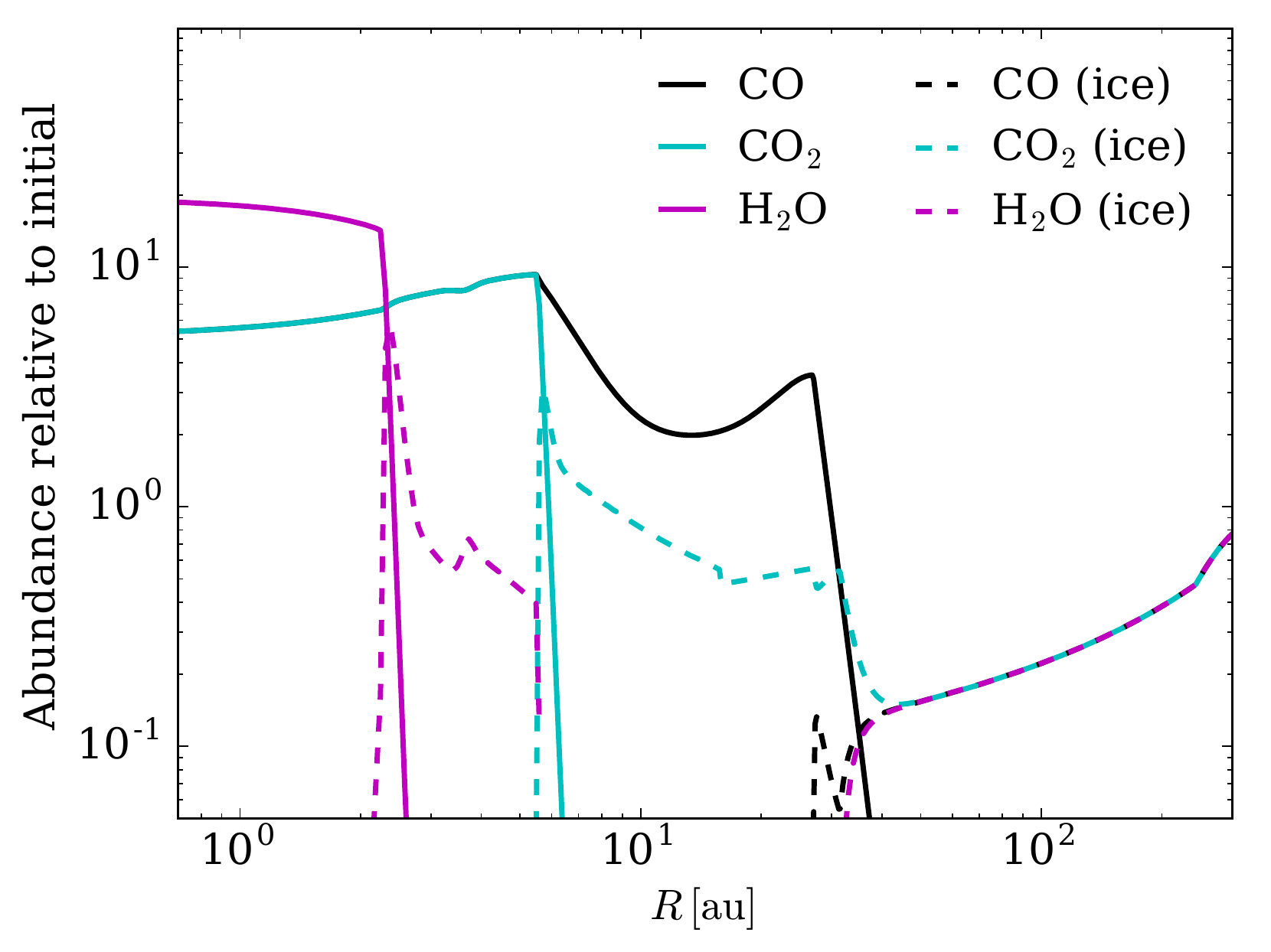}
\end{tabular}
\caption{Abundance of the main carbon and oxygen carriers  relative to their initial abundance at $0.5\unit{Myr}$. These abundances are calculated using the Case 2 chemical model under the assumption  that chemical reactions can be neglected (left) or that equilibrium abundances are maintained(right). In right hand panel the lines showing the relative abundance of CO and CO$_2$ in the gas lie on top of each other inside of the CO$_2$ snow line at 5 to $6\unit{au}$. }
\label{Fig:Molecules}
\end{figure*}

The radial drift of grains from the outer disc carries CO, CO$_2$ and H$_2$O ices towards the star. As grains cross their corresponding snow lines the volatiles evaporate. Since the grains drift faster than the volatiles diffuse, this produces local spikes in the concentrations of molecular species near their snow lines. Inside the CO snow line CO$_2$ and H$_2$O continue to be carried towards the star by radial drift until the molecules meet their own snow lines. Thus the gas becomes enriched in CO just inside the CO snow line, but becomes depleted in CO$_2$ and H$_2$O as they are carried into the inner disc along with the dust grains. This results in the carbon and oxygen abundance becoming dominated by the gas phase CO inside the CO snow line, CO$_2$ inside the CO$_2$ snow line and H$_2$O close to the star. Hence between the CO and CO$_2$ snow line the disc ends up with high C/H ratios and $\mathrm{C/O} = 1$. Closer to the star the C/O ratio decreases because the C/O ratio of the volatiles released at the subsequent ice lines decreases, with the C/O ratio becoming sub-solar in the inner disc despite the high C/H ratio. As a consequence of the radial transport, the disc beyond the CO snow line becomes sub-solar in composition. 

Eventually the flux of dust grains and ices from the outer disc decreases and the disc's evolution is then dominated by the viscous evolution of the gas and diffusion of the molecular species. Since the disc evolves most rapidly at small radii, the water rich gas in the inner disc is accreted before the CO and CO$_2$ rich gas further out. Thus as the disc continues to evolve the C/O ratio in the inner disc increases as the disc becomes increasingly dominated by CO.

As a further illustration, \autoref{Fig:Molecules} shows the abundances of the gas and ice phase molecular species at $0.5\unit{Myr}$ relative to their initial total (gas + ice) abundance. This figure demonstrates that typically a single species is responsible for the gas phase abundance, except inside the water snow line where both the abundance of both H$_2$O and CO$_2$ is high. We also see that the ice composition of the dust grains only changes close to snow lines, with the change in abundance relative to the initial abundance arising due to the change in dust-to-gas ratio.

\subsection{Fast reactions: equilibrium abundances}

\begin{figure*}
 \includegraphics[width=\textwidth]{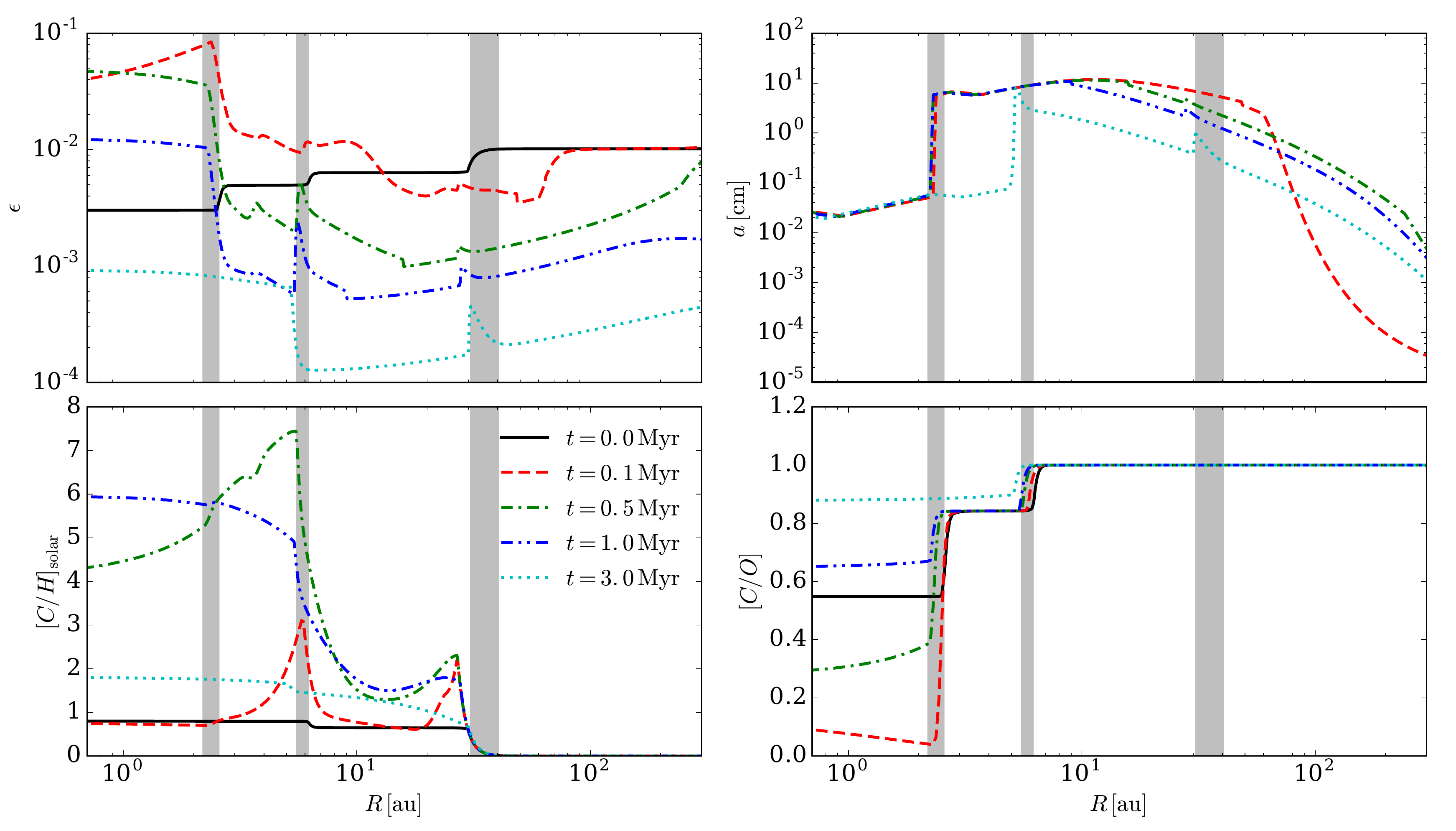}
\caption{Same as \autoref{Fig:ObergDiscTD}, except for the Case 2 chemical equilibrium model.}
\label{Fig:ObergDiscEQ}
\end{figure*}

\begin{figure*}
 \includegraphics[width=\textwidth]{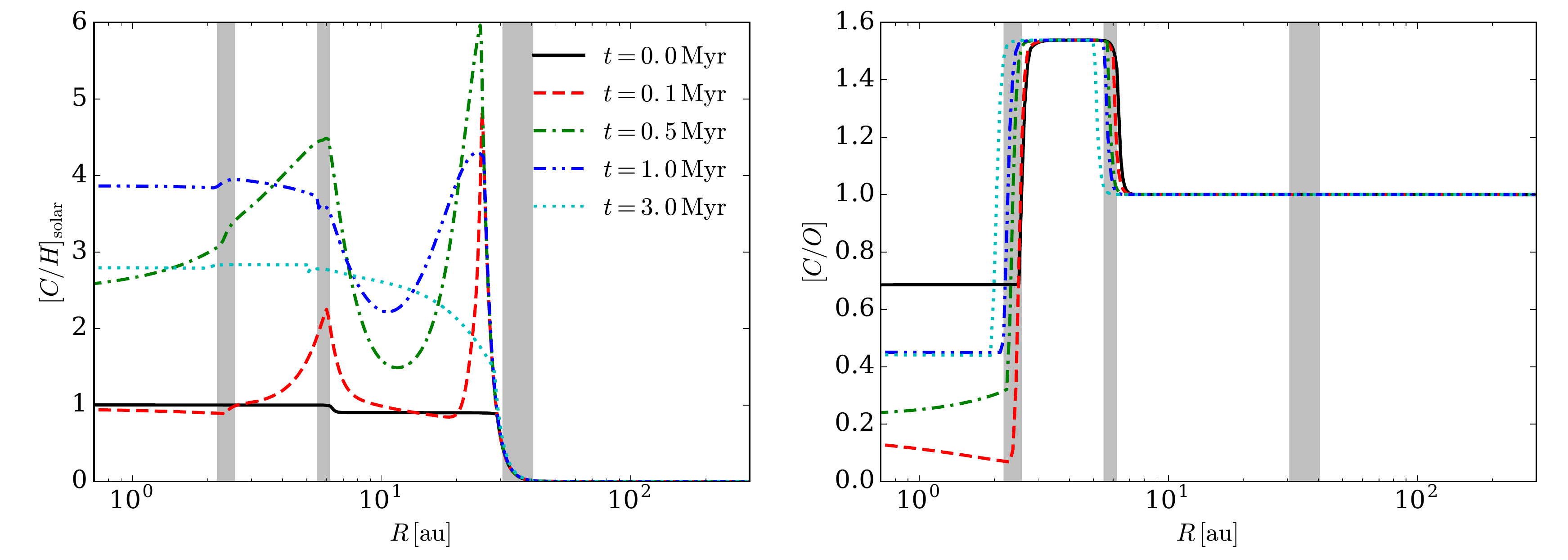}
\caption{Evolution of the C/O and C/H ratios for the the Case 1 chemistry. The evolution of the dust fraction and grain size is similar to the Case 2 chemical models, shown in \autoref{Fig:ObergDiscTD} and \autoref{Fig:ObergDiscEQ}.}
\label{Fig:MadhuChemEvo}
\end{figure*}

The model in the previous section examines how the radial drift of dust grains can redistribute volatiles in the disc when chemical processes occur much more slowly than radial drift. In that case the volatiles are transported rapidly to their ice lines, where they evaporate and are left behind. This leads to large concentrations of particular species close to the ice lines and thus chemical distributions that are out of equilibrium. 

These large imbalances in chemical species may drive chemical reactions. If these reactions occur rapidly enough they may drive the chemical abundances back to a new equilibrium. We investigate the effects that such fast reactions may have on the redistribution of volatiles in a model where the molecular abundances are set back to their equilibrium values based on \autoref{Tab:Chem}. We note that this neglects that different equilibria are possible at different temperatures and compositions; however, it can demonstrate how chemical reactions may affect the redistribution process.

In \autoref{Fig:Molecules} we show the molecular abundances of this equilibrium chemistry model, while \autoref{Fig:ObergDiscEQ} shows the evolution. Although the grain size and dust-to-gas ratio evolution is nearly identical to the slow reaction model, the chemical abundances differ significantly. Notably, the carbon abundance is decreased near the CO snow line relative to the no reactions model, while it is increased inwards from the CO$_2$ snow line. 

\begin{figure*}
 \includegraphics[width=\textwidth]{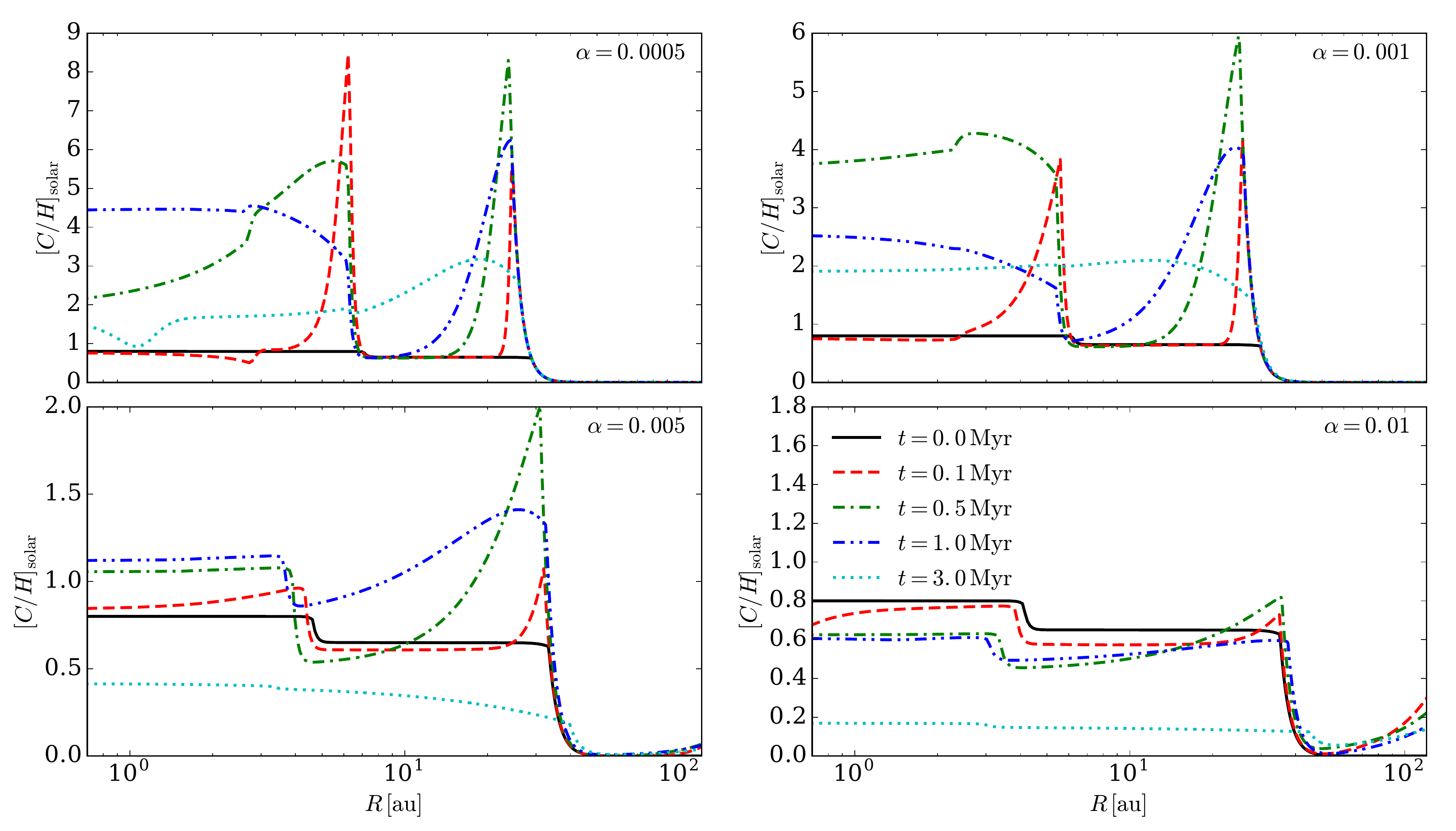}
\caption{Evolution of the carbon abundance for different viscosity parameters, $\alpha$. These models use the Case 2 chemistry ignoring chemical reactions.}
\label{Fig:CarbonCompare}
\end{figure*}

\begin{figure*}
 \includegraphics[width=\textwidth]{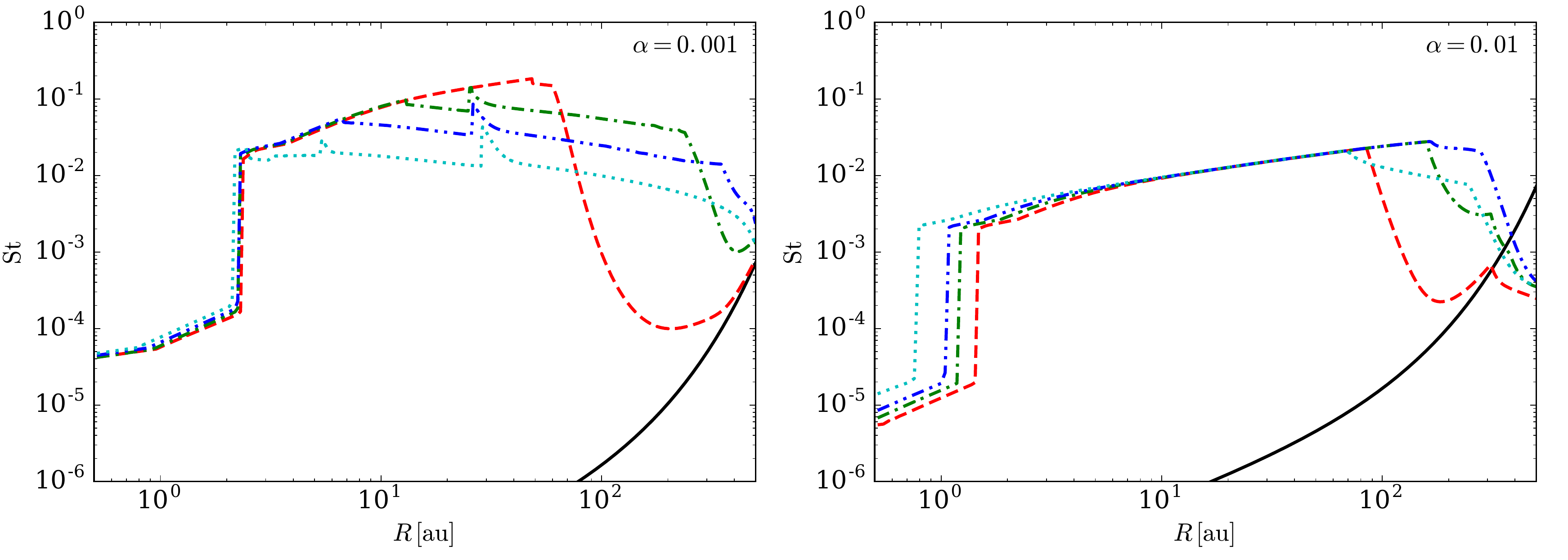}
\caption{Evolution of the Stokes number for two different viscosity parameters, $\alpha=0.001$ (left) and 0.01 (right). The line styles denote the same times as in \autoref{Fig:CarbonCompare}. At $\alpha = 10^{-2}$ grain growth is almost entirely limited by turbulence. This results in a Stokes number that is independent of both the gas density and dust-to-gas ratio (\autoref{Eqn:FragLimit}). At lower viscosity grain growth is instead limited by radial drift in the outer parts of the disc. Since the growth time is proportional to the dust-to-gas ratio this results in lower Stokes numbers (\autoref{Eqn:DriftLimit}). Within about $10\unit{au}$ growth is limited by turbulence even at $\alpha = 10^{-3}$. The smaller Stokes numbers in the outer parts of the disc at early times is because the grains have not yet had time to grow to their maximum size.}

\label{Fig:StokesCompare}
\end{figure*}

\begin{figure*}
 \includegraphics[width=\textwidth]{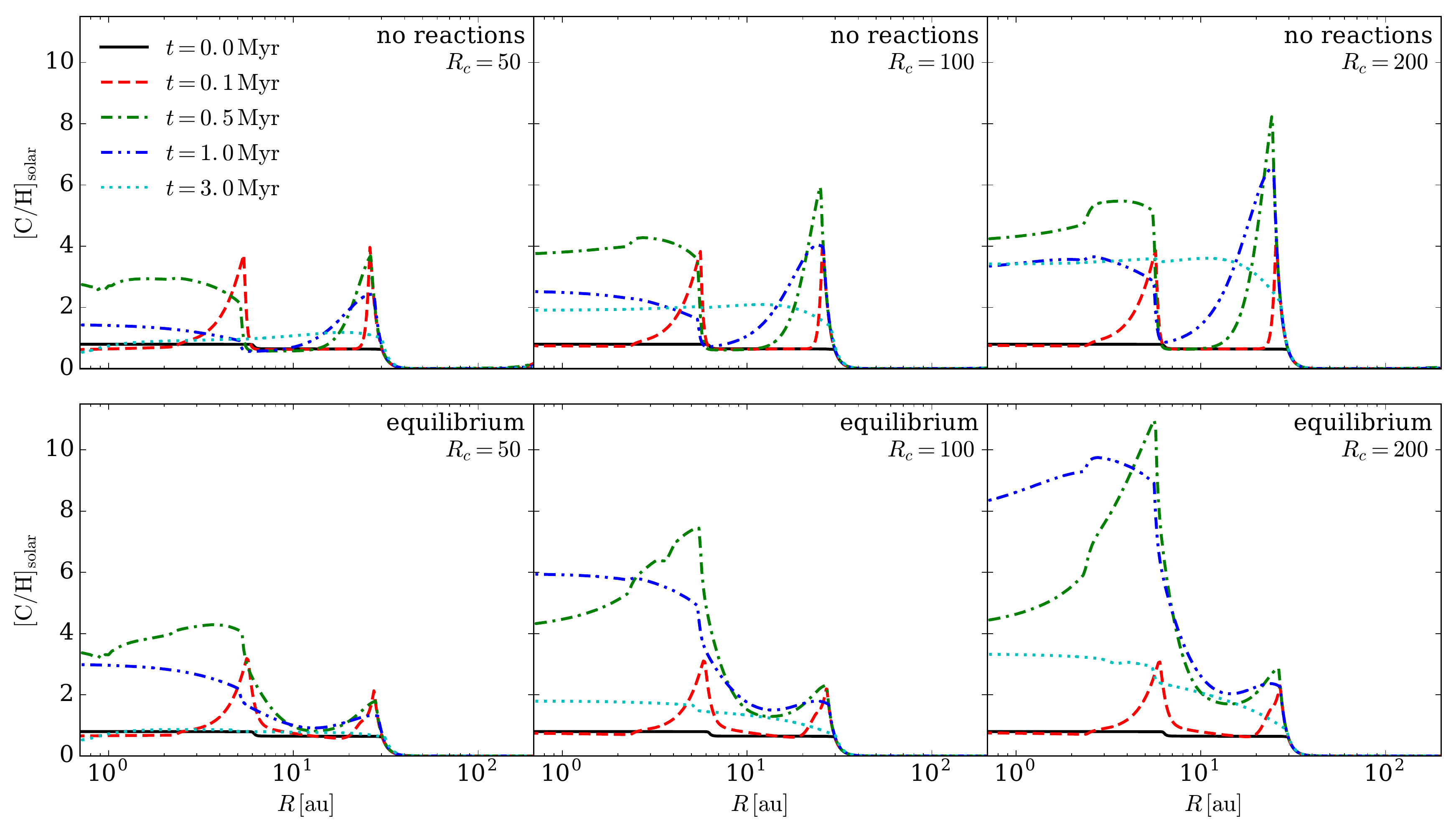}
\caption{Evolution of the C/H ratio for Case 2 chemical models with varying outer radii, $R_c$. The top row shows models in which chemical reactions were neglected, while the bottom row shows models in which they were assumed to be fast enough that the equilibrium abundances were always maintained.}
\label{Fig:RadiusCompare}
\end{figure*}

The higher carbon abundance at the CO$_2$ snow line in the equilibrium model can be understood in terms of an extra transport mechanism for CO. In the no reactions model CO can only move inwards of the CO snow line by  diffusion and gas phase advection. However, in the equilibrium reactions model some of the CO is converted to CO$_2$, reducing the carbon abundance with respect to the no reaction model. This CO$_2$ then freezes out and is transported inwards along with the grains. This produces an over abundance of CO$_2$, some of which is converted back to CO, enhancing the \reply{CO and water ice abundances between the CO and CO$_2$ snow lines.}

\reply{The efficiency of the CO$_2$ ice production and transport depends on the supply of oxygen (in the form of water in our models), required for the conversion of CO into CO$_2$. Within $\sim 10^5\unit{yr}$ the CO abundance becomes high enough at the CO snow line that there is no longer enough oxygen available to maintain the equilibrium between CO and CO$_2$. From this point in time, the flux of CO$_2$ ice is controlled by the supply of oxygen (along with the original CO$_2$) to the CO snow line, which is continually decreasing (along with the dust-to-gas ratio). Thus the CO$_2$ abundance between the CO and CO$_2$ snow lines decreases and the additional CO deposited between the snow lines begins to convert back to CO$_2$. Through this mechanism, the water abundance becomes depleted throughout the  entire region between the CO and CO$_2$ snow lines. At this stage the CO abundance between the snow lines may only continue to change through advection and diffusion in the gas phase, similarly the model in which reactions are neglected.}

By increasing the CO$_2$ abundance on grains this extra mechanism increases the flux of CO$_2$ to the CO$_2$ snow line, increasing the abundance of CO$_2$ there. Although some of the extra CO$_2$ is converted back to CO, no carbon bearing species remain on the grains. Thus inwards of the CO$_2$ snow line carbon bearing species are still only carried to the star by the gas phase advection and diffusion. The higher abundances of CO and CO$_2$ at the CO$_2$ snow line result in higher abundances of these species at all points interior to the CO$_2$ snow line.

In addition to offering an additional way to transport CO inwards of the CO snow line, the imposition of equilibrium abundances allows an transport of CO$_2$ outwards from the CO$_2$ snow line. Normally, the diffusion of CO$_2$ outwards is prevented because any molecules that diffuse beyond their snow line condense onto the grains and are transported rapidly back to the snow line. However, in the equilibrium chemistry model the extra CO at the CO$_2$ snow line is able to diffuse outwards without condensing out, which broadens the snow line outwards. The outwards diffusion of CO can balanced by the conversion of CO into CO$_2$ and subsequently advection on the grains. However the conversion of CO to CO$_2$ is again limited by the availability of oxygen, and thus the carbon enhancement at the CO$_2$ snow line can extend outwards to 10 to $15\unit{au}$.

In the equilibrium models the C/O ratio remains close to its initial value outside of the H$_2$O snow line as the fraction CO and CO$_2$ molecules does not change. Inside of the water snow line, we see that C/O ratio evolves similarly in both models. This shows that despite the disc becoming depleted in water between the CO and CO$_2$ snow lines, radial drift is still delivering water to the inner disc. This occurs due to the conversion of CO$_2$ to CO at the CO$_2$ snow line, freeing up oxygen, which can then be transported inwards as ices.

For the Case 1 chemistry the evolution is largely the same, shown in \autoref{Fig:MadhuChemEvo}. The presence of CH$_4$ inside the CO$_2$ snow line means that the gas phase C/O ratio is higher than in the Case 1 chemistry, just as initial conditions. In both cases the gas phase C/O ratio outside the water snow line is the same as the initial conditions, and thus the same as models in which the chemical composition is taken to be fixed (although the [C/H] ratio is not). The Case 1 chemistry produces a higher C/H ratio just inside the CO snow line than the Case 2 chemistry because a larger fraction of the carbon is in CO. Conversely, the C/H ratio is lower inside the CO$_2$ snow line because the CO$_2$ fraction is lower. The only significant difference between the two models is that inside the water snow line the C/O ratio remains sub solar throughout the simulation because the conversion of CO into CH$_4$ at the CO$_2$ snow line continues to produce water ice long after the water that is initially delivered to the inner disc has been accreted.

\subsection{Viscosity dependence}

Since the source and strength of turbulence responsible for accretion in protoplanetary discs is uncertain, we  examine how these results are affected by changing the turbulent properties of the disc by varying $\alpha$. In these models we have chosen to keep the initial accretion rate fixed, adjusting the initial surface density along with $\alpha$. We note that this choice makes little difference to the results because the radial drift velocity only depends on the Stokes number (\autoref{Eqn:DeltaV}) and the limiting Stokes number in both the radial drift limited and turbulence limited regimes (\autoref{Eqn:DriftLimit} \& \autoref{Eqn:FragLimit}) also does not depend on the surface density normalization. However, the corresponding grain sizes are different. Thus the main effects of varying $\alpha$ are, as $\alpha$ increases: 1) the time-scale on which the gas evolves viscously decreases, 2) the diffusion time-scale decreases and 3) the particle size decreases due to enhanced fragmentation.

We see from \autoref{Fig:CarbonCompare} that for $\alpha \lesssim 5\times 10^{-3}$ the enhancement of the C/H ratio is similar in all models, with the main difference being that stronger diffusion at higher $\alpha$ broadens the spikes at the snow lines. We also see that by $3\times 10^{6}\unit{yr}$ the C/H ratio has dropped below the solar value for $\alpha = 5 \times 10^{-3}$. This is a consequence of the faster viscous evolution at higher $\alpha$. 

At $\alpha = 10^{-2}$ large C/H ratios are never reached. The faster diffusion and viscous evolution is one factor in determining this; however, this is compounded by the delivery of volatiles occurring more slowly than at lower $\alpha$. The slower delivery of volatiles is due to grain growth in the $\alpha = 10^{-2}$ model being limited by high velocity turbulent collisions rather than radial drift, resulting in a lower Stokes number (\autoref{Fig:StokesCompare}). For models with lower $\alpha$, grain growth is limited by radial drift outside of 5 to 10 au instead. Since the radial drift limit does not depend on the level of turbulence or the surface density, in the drift dominated regime the delivery of volatiles to the CO snow line is essentially identical. We note that the turbulent fragmentation also explains the disappearance of the spike in the [C/H] ratio at the CO$_2$ snow line for $\alpha = 5 \times 10^{-3}$: the maximum size of grains transitions from radial drift limited to turbulence limited between the CO$_2$ and CO snow lines in this model, reducing the flux of grains to the CO$_2$ snow line. 

These models also allow us to see how the results would change if the growth time-scale for dust was longer, e.g. due to bouncing preventing growth at large sizes. Slower growth means that radial drift will remove grains before they can become as large, thus reducing the Stokes number and also reducing the flux of ices into the inner disc. Diffusion and gas phase advection would thus be more effective at broadening the peaks near the snow lines. Therefore, a longer growth time-scale has a similar effect to increasing the turbulence. We note that the effect of slower growth will be less dramatic since increasing $\alpha$ affects both the radial drift and diffusion, whilst slower growth only significantly affects the radial drift.

\subsection{Disc size dependence}

\autoref{Fig:RadiusCompare} shows how the size of the discs can affect the abundance evolution. In these models the outer radius of the disc $R_c$ has been varied, with all other parameters fixed. Since the disc mass is proportional to $R_c$ this means that the mass of the discs in these models varies. However, as noted above the evolution is only very weakly sensitive to the normalization of the surface density and thus the difference between these models is due to the different outer radii rather than different disc masses. 

Considering first the model in which chemical reactions are assumed to be slow, we see that models with larger radii have the stronger C/H enhancements after $0.5\unit{Myr}$, but prior to this time the abundance patterns are similar. This can be understood because growth time and radial drift time scales are longer at larger and larger radii, so the pebble flux continues for longer in larger discs.

In the equilibrium chemistry model the delivery of ices to the CO snow line behaves the same way; however, the conversion of CO to CO$_2$ prevents the build up of a large C/H ratio at the CO snow line. Instead, the extra flux of CO and CO$_2$ ices are deposited at the CO$_2$ snow line, resulting in an even larger C/H ratio within $7\unit{au}$.

\section{Planet abundances}

\begin{figure*}
 \includegraphics[width=\textwidth]{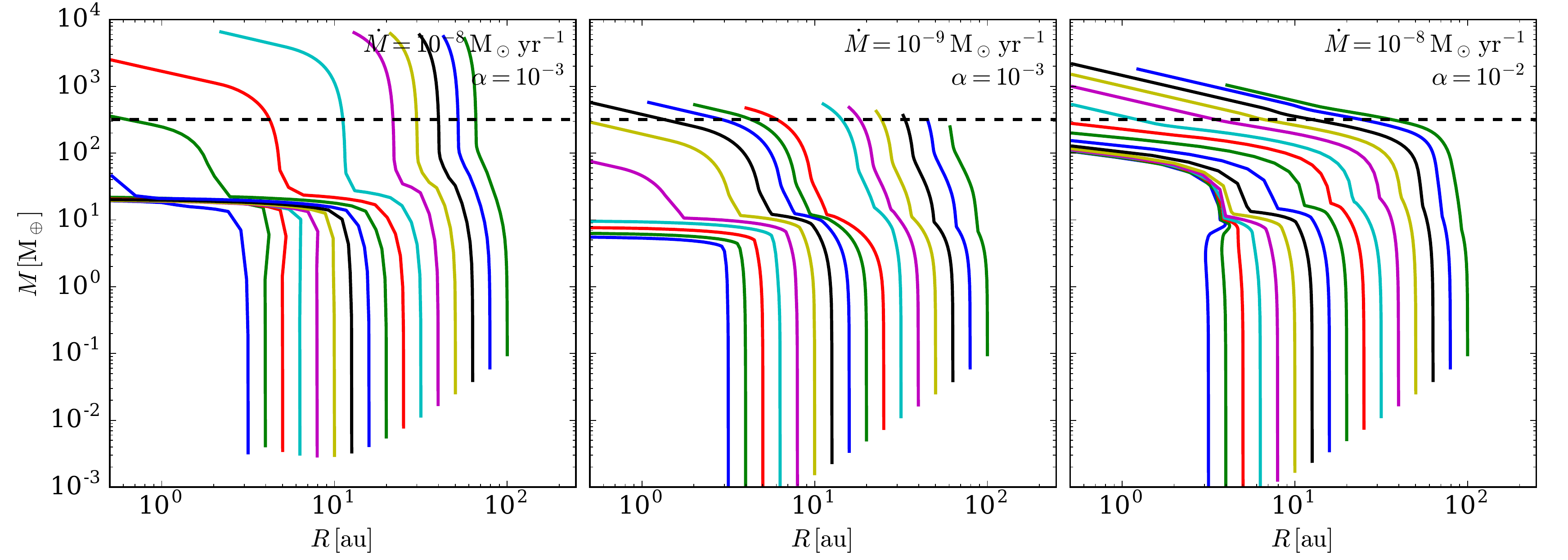}
\caption{Evolution of planet masses and radii up to $3\unit{Myr}$ for different disc models. Each line shows a single track for a planets as it grows and migrates. The tracks shown are for planets forming at start of the simulation, but planets that form later undergo similar evolution but only reach lower masses by $3\unit{Myr}$. The dashed black lines show the mass of Jupiter.}
\label{Fig:PlanetEvo}
\end{figure*}

\begin{figure*}
\includegraphics[width=\textwidth]{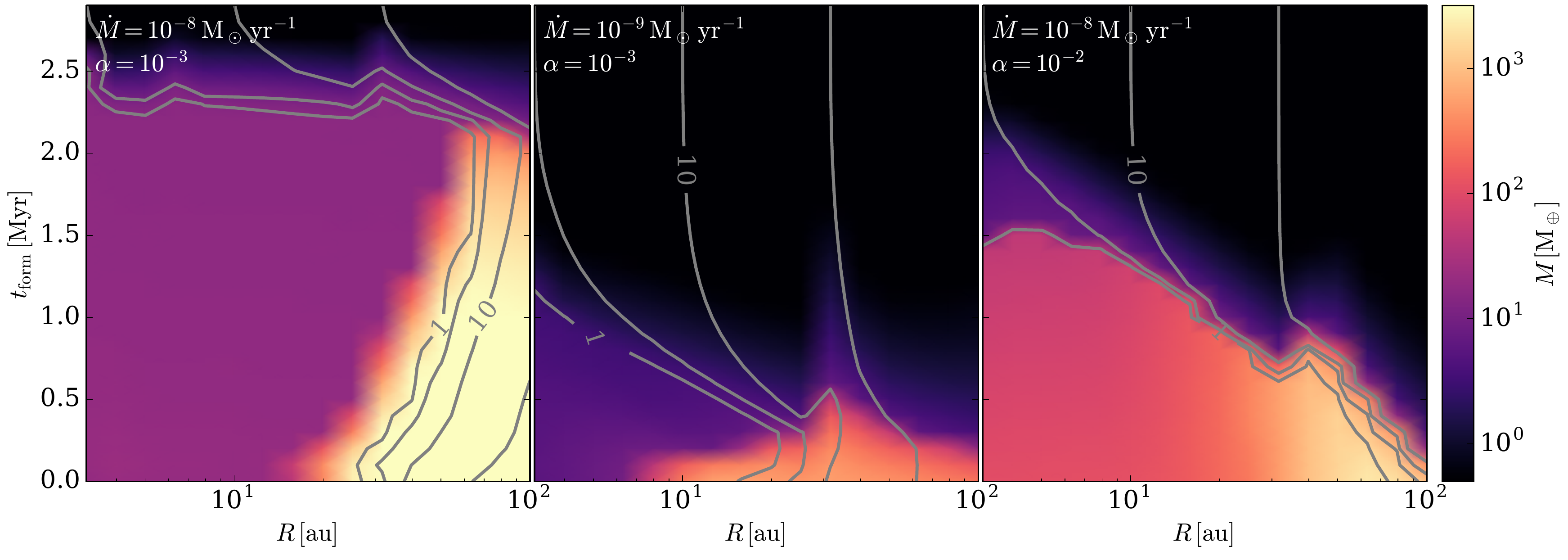} 
\caption{Final masses and locations of planets formed in models with different accretion rates and turbulent viscosities. The $x$ and $y$ axis show the initial formation time and location of the planetary embryo. The colour shows the final planet mass at $3\unit{Myr}$ and the contours show the final location, in $\unit{au}$.}

\label{Fig:PlanetForm}
\end{figure*}

\begin{figure*}
\begin{tabular}{cc}
 \includegraphics[width=\columnwidth]{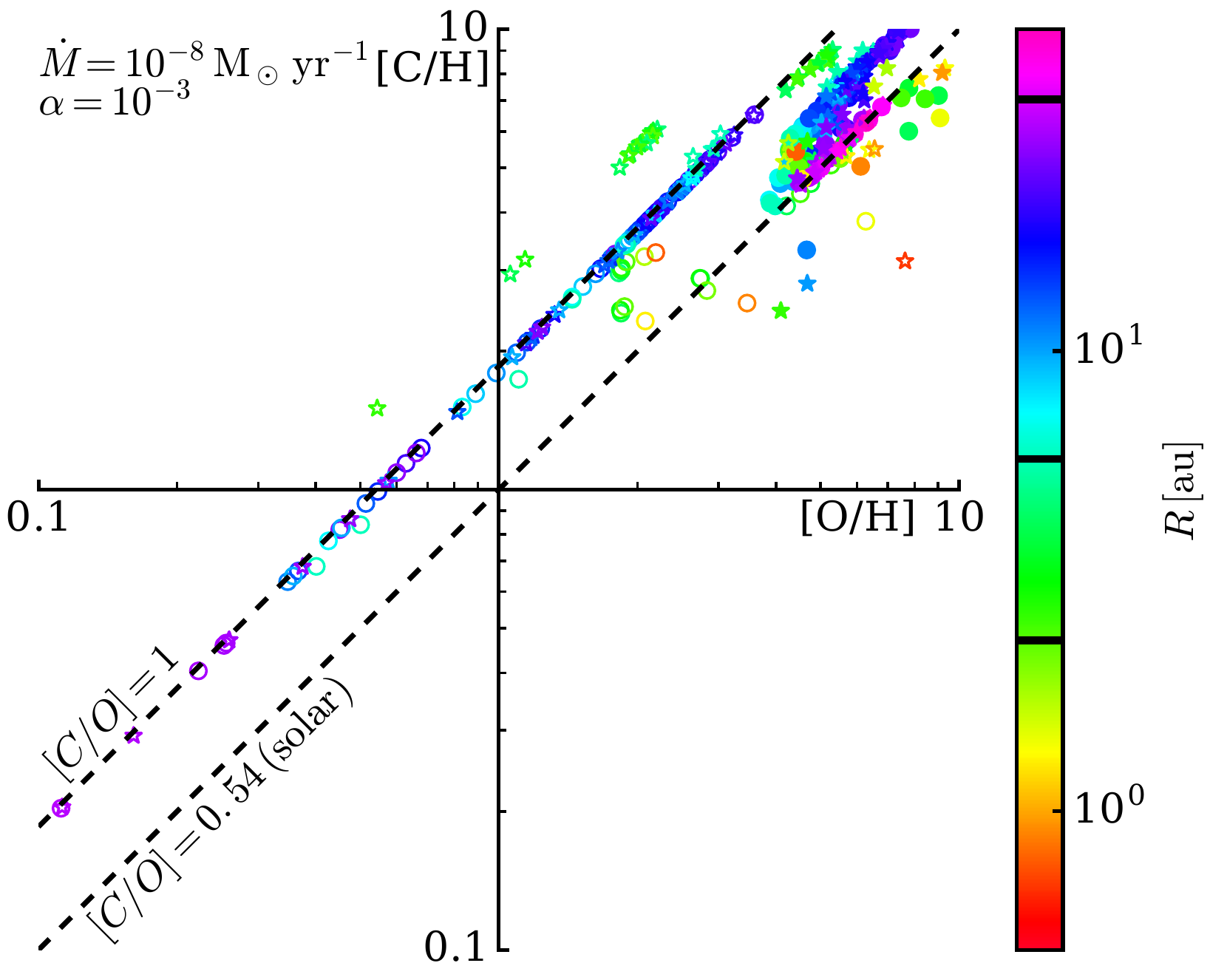} &
 \includegraphics[width=\columnwidth]{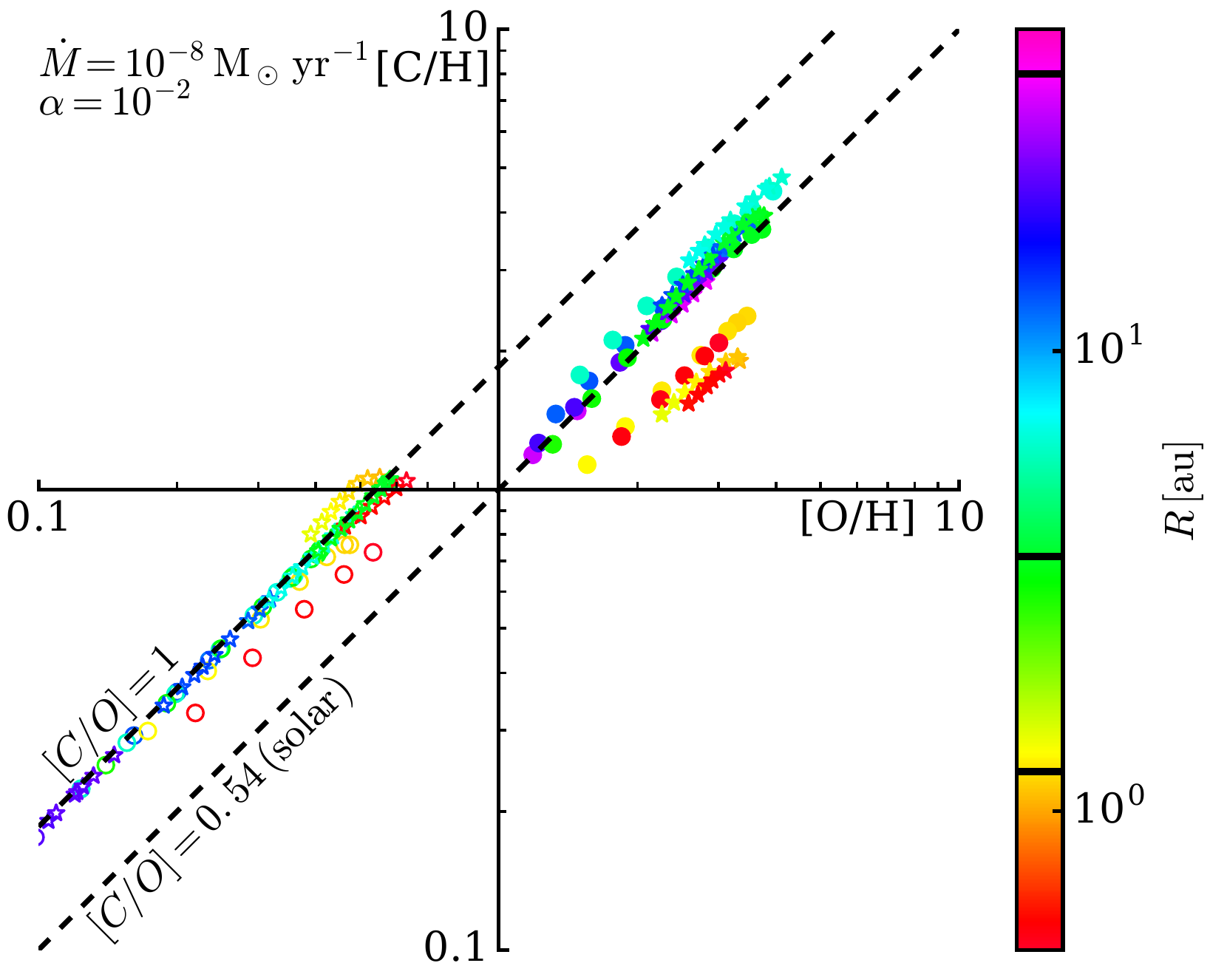} 
\end{tabular}
\caption{Carbon and oxygen abundances of planets at $1\unit{M_{\rm J}}$, formed in a disc with $R_c=200$ and an initial $\dot{M} = 10^{-8}\unit{M_\odot\,yr}^{-1}$. Left: the standard model with $\alpha = 10^{-3}$. Right: High viscosity model with $\alpha = 10^{-2}$. The abundances are given relative to the solar abundances. Stars correspond to the Case 1 equilibrium abundances chemical model, while circles correspond to the Case 2 non-equilibrium abundances. Hollow symbols denote the abundance of the gaseous envelope, while filled symbols denote the total abundance of both the envelope and core. The colour denotes the location of the planets when they reach 1$\unit{M_{\rm J}}$ and the horizontal black lines on the colour bar denote the locations of the water, CO$_2$ and CO snow line (from inside to out). See \autoref{Sec:Results:PlanetChem} for an explanation.}
\label{Fig:PlanetChem}
\end{figure*}

\subsection{Formation \& migration}
\label{Sec:Results:PlanetForm}

The formation and evolution of planets via pebble accretion has been studied in detail in previous studies \citep[e.g][]{Lambrechts2014a,Bitsch2015}, so we only discuss the essential features here. We shall focus on the formation of giant planets, which are relevant for Jupiter, Saturn and giant exoplanets. From \autoref{Fig:PlanetEvo} we see that giant planets tend to start their growth in the outer parts of the disc and migrate inwards. Within a critical radius (a few au to a few 10s of au), the planets migrate too rapidly towards their star through Type I migration for them to accrete massive gaseous envelopes.

The precise details of evolution of the planet populations are model dependent. The most important parameters are the surface density of the disc and the strength of the turbulent viscosity. Since the size of the disc affects how long the pebble flux continues, it can also affect how long giant planet formation occurs at a given location in the disc\footnote{In the inner disc the temperature structure can also be important for controlling the location of planet traps. Since the cores of giant planets tend to form further out, the location of these planet traps are less relevant for our study.}. We explore planet formation in models with $R_c = 200$ for this reason, i.e. in order that giant planet formation may occur over a longer time. We investigate the model dependence using three models, two with initial accretion rates of $10^{-8}\unit{M_\odot\,yr}^{-1}$ but with different $\alpha$ ($10^{-3}$ and $10^{-2}$), and therefore different surface densities. In the final model, we use $10^{-9}\unit{M_\odot\,yr}^{-1}$ and $\alpha=10^{-3}$, which gives the same surface density as the second model. 

\autoref{Fig:PlanetForm} shows considerable differences in the final location and masses of the planets formed in these models. First, we see that the second and third models, which have lower surface densities, can produce giant planets further in. Second, we see that increasing $\alpha$ at fixed surface density allows the planets to become more massive. Third, in the outer disc the growth of planets can be limited by the long growth time-scale associated with a low density of pebbles.

The innermost radius at which planets can accrete massive gaseous envelopes is controlled by the competition between growth and migration. When the pebble accretion rate is high enough, accretion during the envelope contraction phase dominates the time that the planet takes to grow before entering the runaway accretion phase. Since the accretion rate during envelope contraction does not depend on the surface density, but the migration rate is proportional to the surface density, this means that planets reach runaway accretion after less migration at larger distances, or in lower mass discs. In the two models with $\alpha = 10^{-3}$, this explains why the model with the lower accretion rate (and therefore surface density) is able to form giant planets at smaller radii (at $t=0\unit{Myr}$): they have a longer time in which to grow via envelope contraction. However, once the discs have evolved the surface density of pebbles in the lower mass disc becomes too low and the planetary embryos do not accrete gaseous envelopes within $3\unit{Myr}$. 

In addition to the competition between envelope contraction and migration, a second effect is important for determining the outcome of planet formation. This effect is illustrated by comparing the two models with the same initial surface density, but different $\alpha$ and $\dot{M}$. Here we see that increasing $\alpha$ leads to more massive planets, despite stronger turbulence limiting the particle size. The smaller particle size means that the pebble accretion rate at a given surface density is lower, but the radial drift of particles is also slower. Thus in the outer disc the cores become more massive before the surface density of pebbles drops. The larger core mass allows for faster gas accretion during the envelope contraction phase resulting in larger planet masses. 

If we focus on the origin of giant planets, we see that they form  from embryos that typically originate from far outside the water snow line. Neptune to Saturn mass planets may form further in, depending on model parameters. However, we note that when it comes to assessing the chemical abundance patterns of giant planets formed by pebble accretion, the differences between the models make only a moderate difference. As \autoref{Fig:PlanetEvo} shows, we find that giant planets may accrete their envelopes at a range of locations within the disc. In high density environments, migration can prevent planets from accreting their envelopes within water snow line ($\sim 3\unit{au}$). However, we see that planets which end as hot Jupiters can accrete a significant amount of their mass inside the water snow line. In our models the giant planets always continue to accrete gas as they migrate; however, this may not be the case in reality. This caveat, along with the possibility of continued migration after the disc has dispersed (through dynamical interactions or the Kozai-Lidov mechanism, \citealt{Kozai1962,Lidov1962}) means that it may be possible to form hot Jupiters that have accreted their envelopes outside of the water snow line. The converse is not true however: i.e. planets that end up outside the water snow-line are unlikely to have accreted their envelopes within the water snow-line because disc-driven migration is always inwards (unless planet-planet scattering occurred, e.g. \citealt{Lega2013}).

\subsection{Chemical abundances}
\label{Sec:Results:PlanetChem}

We now consider the chemical abundance patterns of giant planets that form in our chemically evolving disc models. While the details of the planet formation process such as the final mass and location of the planets is 
dependent on the planet formation prescription, the chemistry of the planets formed is much less so. Therefore, we focus on the abundances of planets forming in the standard model with $R_c = 200\unit{au}$, $\dot{M} = 10^{-8}\unit{M_\odot\,yr}^{-1}$ and $\alpha = 10^{-3}$. The one exception to this case is when grain growth is suppressed, for example due to strong turbulence driving fragmentation. In this case the radial drift occurs more slowly resulting in a smaller enrichment of the disc. Taking this to the extreme, when radial drift becomes negligible the chemical abundances remain close to their initial values.

The C/H and O/H ratios together provide a useful way of determining how and where the planet formed. They key to this is that inside the CO snow line, the grains always have a C/O ratio that is sub-solar ($\mathrm{C/O} = 0.54$). Conversely, the gas has C/O ratios that are above 0.5, except inside the water snow line. Thus C/O ratios above solar can be used to infer that carbon was predominantly accreted as gas, rather than solids. In this section we demonstrate how the C/O ratio combined with the C/H ratio can be used to infer the location in which the planet formed. To this end, we place the planets formed in our models on this diagram in \autoref{Fig:PlanetChem}, when the planets reach the mass of Jupiter. 

When the gas is not significantly enriched by radial drift, we find that only one of the C/H and C/O ratios may be significantly super-solar. Similar abundance patterns were found in previous models that neglected radial drift \citep[e.g.][]{Madhusudhan2014b, Madhusudhan2016}. One such case is the model with $\alpha = 10^{-2}$, as shown in \autoref{Fig:PlanetChem}. Here the accretion of gas alone produces planets with C/H ratios that are sub-solar, but C/O ratios between solar and C/O = 1, independently of where they form. Therefore, in order to produce super-solar C/H ratios some enrichment by solids is needed, either through  erosion of the core, or enrichment by planetesimals. Typically this produces a sub-solar C/O ratio because the ices are oxygen rich. Only outside the CO snow line does the C/O ratio in the ices become solar. Pebble accretion in such environments can produce a \emph{slightly} super-solar C/O ratio because the cores of giant planets may migrate a long way before accreting their gas (see \autoref{Fig:PlanetEvo}). Thus planets can form cores outside the CO snow line, which have solar C/O ratios and then accrete their gas inside the CO snow line, where the gas has $\mathrm{C/O} = 1$. However, while this may produce C/O ratios above solar, this will never produce a C/O ratio close to unity, unless the the C/H ratio remains sub-solar.

Conversely, the envelopes of giant planets may have C/H and C/O ratios that are both super-solar if they accrete metal rich gas. This is made possible by the radial drift of large grains, which carry volatile species to their ice lines, enriching the gas when they sublimate. These volatiles may then spread inwards, enriching the entire region inside of their snow lines. In what follows, we first consider the compositions of gaseous envelopes formed by accreting metal rich gas, \emph{without} any enrichment by solids. Subsequently we discuss the effects of enrichment.

The C/O ratio of planets that accrete their envelope outside of the CO$_2$ snow line is always 1 (neglecting enrichment by solids). This is because CO is the only carbon or oxygen bearing species in the gas phase. However, due to radial drift the C/H ratio of planets with $\mathrm{C/O} = 1$ varies considerably due to the large variation in the C/H of the disc, both radially and temporarily. Between the CO and CO$_2$ snow line the C/H ratio of the disc is super-solar, and thus planets that accrete their envelopes here also have super-solar C/H ratios. Outside the CO snow line the gas phase C/H ratio is negligibly small, so planets that accrete their entire envelopes outside the CO snow line would be off the bottom left of \autoref{Fig:PlanetChem}. Those planets that do have $\mathrm{C/O} = 1$ and C/H of solar or slightly below can be formed in two ways. Either they accreted part of their envelopes in the CO depleted gas from outside the snow line along with some CO rich gas inside the snow line, or they accreted their envelopes late in the disc's evolution once the global C/H ratio has dropped.

Planets that accrete their envelopes inside the CO$_2$ snow line have $\mathrm{C/O} \neq 1$ because CO$_2$ and CH$_4$ also contribute to the carbon and oxygen abundance. The C/O ratio differs between the models due to the different amounts of these species present, but all models predict $\mathrm{C/O} \neq 1$. If chemical reaction occur more slowly than radial drift, the gas between the CO$_2$ and H$_2$O snow lines is dominated by CO$_2$ so we find  $\mathrm{C/O} \approx 0.5$ to 0.6, i.e. close to solar. If instead chemical reactions are fast enough to maintain chemical equilibrium between the carbon and oxygen bearing species, the C/O ratio remains close to its initial value in this region, i.e $C/O \approx 1.5$ and 0.85 for the Case 1 and 2 models respectively. The C/H ratio can vary considerably depending on the exact location and model, but the C/H ratio between the CO$_2$ and H$_2$O snows lines is generally super-solar when radial drift is efficient.

Sub-solar C/O ratios are only produced by gas accretion when planets accrete their envelopes inside the H$_2$O snow line. Even in this region we typically find that radial drift produces C/H ratios that are super-solar and thus there is a region of the C/H and O/H parameter space that does not contain any planets (where the O/H ratio above solar and the C/H ratio is less than about twice solar). The precise size of this region is again likely to be model dependent, but similarly to models that neglect radial drift, planets do not form with  super solar O/H and sub-solar C/H. \autoref{Fig:PlanetChem} contains relatively few planets in this region because many of the planets that accrete a significant fraction of their envelope inside the H$_2$O snow line end up with masses of several Jupiter masses. For this reason, hot Jupiters that have a super-solar C/H ratio and a sub-solar C/O ratio may be more frequent than the density of points in \autoref{Fig:PlanetChem} suggests.

The accretion of chemically enriched gas leaves clear signatures in the abundances of giant planets. This can be recognised through planets that have C/O and C/H ratios that are both higher than that of their host star, since these abundances cannot be reached when enriching the atmosphere with solids. Furthermore, because the gas is dominated by CO beyond the CO$_2$ snow line, accurate measurements of the C/O ratio will be able to determine whether the planet formed outside the CO$_2$ snow line or not. 

In the case of metal enrichment via solids, it may be possible to determine whether these solids were accreted through planetesimals, or partially dissolving the core. This is possible because C/O ratios above solar only appear if the solids were accreted outside the CO snow line, but the gas was accreted inside it. Since enrichment via planetesimal accretion cannot happen before the envelope is accreted, this cannot occur without invoking outward migration. Finally, the combined gas of both metal enriched gas accretion and subsequent enrichment by solids should be identifiable through a population of planets with very high C/H ratios ($\gtrsim 5$) and C/O ratios between roughly solar and $\mathrm{C/O} = 1$.

\subsection{Jupiter \& Saturn}
\label{Sec:Results:Jupiter}

In comparison to exoplanets, the chemical abundances of the solar system giants are uniquely well determined. Jupiter is known to be metal rich, with a C/H ratio that is 3 to 5 times solar \citep{Owen1999}, along with super-solar abundances of nitrogen, sulphur and noble gases. However, the water abundance is still uncertain. Observations of a hot spot made by the Galileo probe suggest a vastly sub-solar O/H ratio, but this may not be representative of Jupiter's bulk abundance. On the other hand thermo-chemical modelling of Jupiter's structure suggests the water abundance should not be enhanced more than about 8 times solar \citep{Visscher2010}, and thus the C/O ratio should not be lower than about half of the solar value. For Saturn, the abundance pattern appears similar, but with a C/H ratio of about 8 times solar \citep{Atreya2016}. Although the C/O ratio of the solar system giants is currently unknown, Juno will soon measure Jupiter's water abundance \citep{Matousek2007}, which can be used to constrain models of planet formation.

Planet formation models constrain Jupiter and Saturn to be somewhere close to its current location during the terrestrial planet formation era \citep[and references therein]{Raymond2014}, ruling out disc-free migration modes for them. Therefore, we know that they likely accreted their envelopes somewhere near their current locations, unless gas accretion can be stopped without stifling migration.

At $\sim5\unit{au}$, Jupiter likely accreted its gaseous envelope somewhere close to or inside the CO$_2$ snow line, which is at 7\unit{au} in our models. For models that produce a Jupiter mass planet close to its current location we find that the equilibrium chemistry for both Case 1 and 2 are easily able to produce C/H ratios in the envelope that are 4 to 5 times solar. However, if the reactions are slow then the delivery of CO molecules to the CO$_2$ snow line is slower and the C/H ratio is instead 2.5 to 3.5 times solar, compatible with estimates at the lower end of the range for Jupiter. 

The C/O ratio of our Jupiter models depends on which side of the CO$_2$ snow line Jupiter accreted its mass. If Jupiter accreted its envelope outside the CO$_2$ snow line then it should  have $\mathrm{C/O} \approx 1$. While different chemical models make differing predictions for the C/O ratio if Jupiter accreted its envelope inside the CO$_2$ snow line, in general we find the C/O ratio differs from  $\mathrm{C/O} \approx 1$. The Case 2 no reaction model produces $\mathrm{C/O} \approx 0.5$ because the carbon abundance is dominated by CO$_2$ inside the CO$_2$ snow line in this case. For the equilibrium models, the C/O ratio remains close to the initial values (\autoref{Fig:ICs}), producing a C/O ratio of 1.5 and 0.8 for Cases 1 and 2 respectively. We also find models that accrete similar amounts of mass on both sides of the snow lines, in which case the C/O ratio can be intermediate between these values. 

Forming Saturn without solid accretion is more problematic because its C/H ratio is observed to be 8 times solar. At $9.5\unit{au}$ it Saturn is outside of the CO$_2$ snow line, but also a long way from the CO snow line. Even in the most optimistic of our models Saturn would need to have accreted its envelope very close to the CO snow line to produce a sufficiently high C/H ratio. For this reason the enrichment of Saturn's atmosphere by solids is appealing. From such models we expect $[{\rm C/O}] \approx 0.6$ to 0.7. 

We note that it might be surprising to find that the metals in Jupiter's atmosphere were accreted as metal rich gases, but in Saturn they arrived in solids. Thus  one might expect Juno to find Jupiter has a significantly sub-solar C/O ratio. A second possibility would be that Saturn accreted its envelope inwards of its current location, close to the CO$_2$ snow line, and migrated outwards to its current location. Such outward migration is proposed by the grand-tack model of terrestrial planet formation \citep{Walsh2011}. However, some enrichment by solids is likely required, otherwise highly favourable conditions for gas enrichment such as low turbulence ($\alpha < 10^{-3}$) in the inner disc or a large outer radius ($\gtrsim 200\unit{au}$) would still be needed in order to produce high enough C/H ratios.

\section{Discussion}
\label{Sec:Discuss}

\subsection{Disc evolution}

In this work we have investigated the redistribution of the dominant carbon and oxygen bearing species in evolving protoplanetary discs, along with the associated impact on the chemical abundances of giant planets that form in such discs. The radial drift of icy grains from the outer disc can supply volatiles to the inner disc on shorter time-scales than they can be carried away by gas phase diffusion or advection, resulting in the enhancement of volatiles at the snows lines. As the disc evolves, these enhancements spread inwards, enriching the gas phase C/H ratio in the inner disc by factors of a few to several. 

A side effect is the segregation of molecular species in the disc. CO molecules are predominantly left at 20 to $30\unit{au}$, near the CO snow line, while CO$_2$ and water travel further in before being left behind at their own snow lines. Inside the CO and CO$_2$ snow lines this drives the C/O ratio towards that of the dominant species. However,  chemical reactions will play an important role in determining how strong the change in C/O ratio is because they may be able to remove some of the excess molecules. We explored the effect of chemical reactions have on radial transport by considering two limiting cases: 1) the case where they occur on time scales longer than the time scale for disc evolution, which we model by neglecting them entirely 2) the case where the chemical time-scale is fastest time-scale, modelled by resetting the molecular ratios to their initial (`equilibrium') values. In the first model the C/O ratio inside each of the snow lines becomes dominated by their respective molecules (e.g. CO inside the CO snow line), while in the second the C/O ratio stays close to the initial value. In practice, the time scale for chemical reactions lies between these extremes (\citealt{Eistrup2016} estimate a few $10^5\unit{yr}$) and thus the degree to which the local abundances are dominated by the species deposited at each snow line will also fall between the two extremes.

In addition to restoring the initial C/O ratio, rapid chemical reactions also enhance the transport of carbon inside the CO snow line, resulting a greater fraction of the carbon bearing species being deposited at the CO$_2$ snow line. This also introduces differences into the late phase of the disc's evolution, once the flux of ices has fallen, which is governed by viscous accretion. During this phase the chemical abundances are dominated by the gaseous species, so the evolution can be understood by considering only the gas. Since accretion removes mass sequentially from inside to out, the water rich inner disc is cleared first and is replaced by the carbon rich gas that has moved in from between the water and CO$_2$ snow lines, thus the C/O ratio becomes constant throughout out the region inside of the CO$_2$ snow line. Accretion then removes the gas that was initially inside the CO$_2$ snow line and the C/O ratio tends to unity as the whole disc becomes dominated by the CO rich gas from outside of the CO$_2$ snow line. Since a greater proportion of the carbon bearing molecules are deposited at the CO$_2$ snow line in the case of rapid chemical reactions, the C/H ratio also decreases faster because it is accreted earlier.

At the snow lines we find an enhancement in the dust-to-gas ratio of up to factors of a few. The peak of the solid density occurs just outside of the peak of the vapour concentration. This peak is driven by the outward diffusion and recondensation of ices onto grains that are just outside the snow line. Unsurprisingly, these results agree with \citet{Schoonenberg2017}, who model essentially the same processes: radial drift, diffusion and thermal adsorption and desoprtion. We also see that where the grains transition from being icy to ice free, there is a jump in the surface density, which occurs inside the snow line. In our model this arises because the ice free grains fragment more easily, as was reported by \citet{Cridland2017} who used a similar model for the water snow line. \citet{Schoonenberg2017} also find similar behaviour if the grains have the structure of many small silicate cores held together by ices. In this case sublimation drives fragmentation. Sintering also gives rise to a similar effect \citep{Saito2011}. 

\citet{Stammler2017} have conducted a similar study into the structure of the CO snow line. They neglected fragmentation and instead find a different behaviour, that the growth of grains driven by the extra adsorption of ices outside the snow line leads to larger grains that drift faster. Neglecting fragmentation at the CO snow line is likely a valid assumption, because the collision velocities are lower and the water ices remain on the grains and so collisional fragmentation is not expected. However, at the water snow line there is likely to be a complex interplay between growth and fragmentation. Whether an enhancement in the dust-to-gas ratio should be expected at the water snow line will depend on the details of grain growth and fragmentation. However, these differences are unlikely to have much of an effect on the evolution of the chemical abundances, because the mass trapped outside the snow line is a relatively small fraction of the volatile abundance inside the snow line. 

Our suggestion that radial drift should lead to chemical enrichment in the inner disc is in tension with observational constraints. For example, thermochemical modelling of HD, CO and carbon lines in TW Hydrae suggests that carbon should be depleted by perhaps as much as a factor of 100, along with a C/O ratio that might be above unity \citep{Favre2013,Du2015,Kama2016a,Manara2016,Bergin2016}. These observations are somewhat difficult to reconcile with our models, which suggest that CO should be enhanced by radial drift. One possibility is that a greater fraction of the carbon bearing species is locked up in ices than is predicted by  models. This might change the detailed abundances of our models by altering the transport of the volatiles, but the phenomenology should not be affected. \reply{Alternatively, the low C/H ratio can be achieved during the vary late phase of our models, once the majority of the CO and CO$_2$ has been accreted onto the star. In this case the low C/H ratio should be associated with a low dust-to-gas ratio. In many of our models the dust-to-gas ratio drops to a few $10^{−4}$ before the CO in the inner disc is accreted because the dust evolution is much faster than the gas evolution. Therefore in order to achieve low C/H before exhausting the disc of dust it may be necessary to invoke high viscous $\alpha$. This both accelerates the accretion of CO and limits dust depletion: high $\alpha$ promotes dust fragmentation and hence, by limiting particle size, slows down radial drift.  Given TW Hydrae may have a relatively high accretion rate, this may be a reasonable explanation \citep{Brickhouse2012}.}

Another suggested explanation for the depleted C/H is that most of the carbon is in the form of refractory organic molecules. If this explanation is true, then the enhancement of the C/H at the snow lines would not occur as the refractory carbon would remain on the grains. However, there is good observational evidence that refractory species cannot dominate the carbon abundance. In Herbig Ae stars the long mixing time-scale means that the abundance of the stellar atmosphere can be used to measure the abundance of the gas accreted. \citet{Kama2015} showed that in transition discs with large holes the gas accreted is depleted in refractory elements such as Fe, Mg and Si, while non-transition discs  showed no depletion. This can be taken as evidence that the transition objects were accreting gas that is depleted of dust, as would be expected if planets are responsible for opening up gaps in the disc. However, both transition and non-transition discs have stellar carbon and oxygen abundances that are compatible with solar values, which is strong evidence that neither of these species can be predominantly in refractory species on the grains.  One possible solution to this problem is conversion of CO into CO$_2$ ice \citep[e.g][]{Eistrup2016}, in which case we would expect the chemical enrichment  to follow our equilibrium abundance models.

There is also evidence that disc evolution can proceed as we model it here, i.e. with the radial drift of grains occurring on time-scales shorter than the gas evolves. The numerous discs in which the dust radii are observed to be smaller than their gas radii \citep{Guilloteau2011,Andrews2012,Pietu2014,Birnstiel2014} provide evidence for on going radial drift. Furthermore, there are now a number of debris discs known that have low dust masses (such as HD 141569 and HD 21997), but still have large amounts of gas as traced by $^{18}$CO and $^{13}$CO \citep{Matra2015,Wyatt2015,Pericaud2016}. The CO masses derived are too high to be compatible with a cometary origin for CO, suggesting that it must be primordial \citep{Kospal2013,Thi2014}. These objects provide an ideal laboratory to test our models against disc evolution. We should expect to see a high C/H ratio due to the radial flux of ices, and an inner disc that is depleted in water since it will likely have been accreted onto the star. These objects should also have super-solar C/O ratios.

\subsection{Planet formation}

We have examined the carbon and oxygen abundances of planets that form in evolving discs. While there is considerable differences between the evolution tracks of planets forming at the same location in different disc models, the chemical abundances of the planets only depend on where they accreted their envelopes. A key feature of the radial drift of ices is that the gas phase abundances of the dominant carbon and oxygen carriers can become enriched. This means that giant planets with metal rich atmospheres need not have enriched their atmospheres by the accretion of solids (through either planetesimal accretion or a partially dissolved core).

The formation of metal rich atmospheres by the accretion of metal rich gas produces clear signatures in the C/O ratio, which can be used to determine that their abundances came from gas accretion rather than solid accretion. Furthermore, the C/H and C/O ratios can be used to determine \emph{where} in the disc the planets formed in relation to the various snow lines. The explanation for this difference comes from the fact that inside the CO snow line the composition of ices is always oxygen rich (C/O less than solar), while with the exception of inside the H$_2$O snow line the C/O ratio is always close to solar (CO$_2$ dominated gas, $\mathrm{C/O} = 0.5$) or super-solar. 

Since these high C/O and C/H ratios cannot be formed by enrichment of the atmospheres by solids, the discovery of planets with such abundances can be used to constrain the evolution of the discs in which they formed. The formation of planets with a low C/O and high C/H ratio remains possible through enrichment by solids, or accretion of the envelope within the water snow line. According to our models, planets with low C/H ratios must have formed beyond the CO snow line, or late in the disc's evolution once the pebble flux has dropped and the C/H ratio has decreased through accretion onto the star.

By putting upper limits on the amount of solids accreted by giant planets, it may be possible to constrain planet formation models. For example, in the case of a  planet with a precisely determined C/O ratio that is close to unity, along with a high C/H ratio, we can be sure that it accreted its metals through the accretion of metal rich gas from outside the CO$_2$ snow line. It may thus be possible to put an upper limit on the amount of solids accreted and thus limit the surface density of planetesimals in the disc. Given the need for high planetesimal surface densities for planetesimal accretion to be efficient \citep{Kobayashi2011}, along with the fact that a high gas phase C/H ratio implies significant radial drift and the presence of pebbles, the existence of such planets may be able to distinguish giant planet formation by pebble accretion from planetesimal accretion.

Current observations of hot Jupiters are often consistent with either elevated C/O ratios or possibly C/O ratios close to that of their host star as the uncertainties remain significant \citep[e.g.][]{Line2014, Madhusudhan2014b,Kreidberg2015,Brewer2017}. Estimates of the O/H (and thus C/H) ratios for these planets inferred from H$_2$O abundances have produced a wide range of abundances that vary from consistent with solar to highly sub-solar. All current planet formation models require formation outside of the CO snow line to produce such low O/H or C/H ratios, so it is already clear that some planets may have formed far out in the disc. For those with an O/H ratio closer to solar, more accurate C/O and C/H ratios should be able to determine how they accreted their metals. 

For the solar system giants, the uncertainty in their C/O ratio means that we cannot yet be sure how or where they formed. However, Juno will soon make this measurement \citep{Matousek2007}. Our models make clear predictions for Jupiter's C/O ratio based on its formation location. If Jupiter accreted its metals through the accretion of metal rich gas, then we expect $\mathrm{C/O} \ge 0.5$, i.e close to solar and above, rather than sub-solar as would be expected for the accretion of solids. If Juno finds that Jupiter has $\mathrm{C/O} = 1$, then we expect Jupiter to have formed outside the CO$_2$ snow line, while $\mathrm{C/O} > 1$ would imply that Jupiter formed in a methane rich environment inside the CO$_2$ snow line, while $0.5 < \mathrm{C/O} < 1$ also implies formation inside the CO$_2$ snow line, but in a CO$_2$ rich environment.

\section{Conclusions}
\label{Sec:Conclude}

We have investigated the enrichment of the gas phase metallicity in discs and giant planets that is associated with the radial drift of icy grains from the outer parts of the disc. By following the condensation and evaporation of the dominant carbon and oxygen bearing species, CO, CO$_2$, CH$_4$, and H$_2$O, we are able to determine how the C/H, O/H and C/O ratios evolve as the disc evolves. Since the radial drift of dust grains is typically faster than the gas evolution, this leads to enhancements in the metallicity of the disc near the snow lines, where the molecules sublimate. Near the snow lines this can result in enhancements of up to a factor of 10 in the C/H or O/H ratios. These enhancements can then spread inwards as the gas phase molecules diffuse and are carried towards the star along with the gas, enriching the entire disc inside the CO snow line (located at about 30 to $40\unit{au}$). The enhancement of the C/H and O/H ratios in the disc opens up the possibility that giant planets may become metal rich by accreting metal rich gas directly, rather than requiring subsequent enhancement by the accretion of solids (either through the accretion of planetesimals or dissolving part of the planet's core into the envelope). 

The accretion of metal rich gas can produce planets with C/H and C/O ratios that are both super-solar. This region of parameter space cannot be reached by enrichment of the atmosphere due to solids alone, because the C/O ratio of icy dust grains is always sub-solar \citep{Oberg2011}. Thus the detection of planets with C/O and C/H ratios that are both super-solar is clear evidence that the majority of the planet's metals were accreted from metal rich gas. By determining limits on the amount of solids accreted in the form of planetesimals, it may be possible to put upper limits on the surface density of planetesimals present in the disc when the planet accreted its envelope. Given the need for a relatively high surface density of planetesimals for traditional core accretion models to form giant planets \citep{Pollack1996,Kobayashi2011}, it may thus be possible to determine whether such a planet formed by planetesimal accretion or pebble accretion.

Furthermore, the C/O ratio and C/H ratio of the planet can be used to constrain where the planet accreted its envelope. For example, outside the CO$_2$ snow line the gas phase carbon abundance is dominated by CO and thus the C/O ratio is close to 1. Similarly, we find that C/O ratios below about solar are only possible if the planet accreted its envelope inside the H$_2$O snow line, and thus very oxygen rich hot Jupiters would favour in situ formation scenarios. Additionally, the C/H ratio can be used to infer where the planet formed in relation CO snow line, since outside of the CO snow line all of the dominant carbon bearing species are frozen out onto grains, and thus the planets should have C/H ratios that are much smaller than solar. 

A similar argument can be applied to Jupiter to determine where it formed. Although Jupiter's C/O is currently poorly known, the Juno satellite will soon provide a measurement \citep{Matousek2007}. Given Jupiter's current location ($\sim5\unit{au}$) is close to the CO$_2$ snow line (located at $7\unit{au}$ in our models), depending on how far it migrated Jupiter may have accreted its envelope on either side of snow line. If Juno measures a  C/O ratio close to unity, this would be good evidence that Jupiter formed outside of the CO$_2$. Our predictions for Jupiter's abundance if it formed inside the CO$_2$ snow line are dependent on assumptions made in the chemical evolution model, but we expect $\mathrm{C/O}\neq 1$. Since none of our models produce C/O ratios that are below solar, a sub-solar C/O ratio measurement would support the idea that Jupiter's metals were instead delivered by the accretion of solids. 

\section{Materials}

In the interest of reproducibility, the code used to generate the models in this paper has been made freely available and can be found on \href{https://github.com/rbooth200/DiscEvolution/tree/CO_evol_paper}{github}.

\section{Acknowledgements}

\reply{We thank the reviewer for comments that helped improve the clarity of the manuscript.} This work has been supported by the DISCSIM project, grant agreement 341137 funded by the European Research Council under ERC-2013-ADG.

\bibliography{chemo_drift}
\bibliographystyle{mnras_edit}

\bsp

\appendix

\label{lastpage}
\end{document}